\documentclass[fleqn,usenatbib]{mnras}

\usepackage{newtxtext,newtxmath}

\usepackage[T1]{fontenc}

\DeclareRobustCommand{\VAN}[3]{#2}
\let\VANthebibliography\thebibliography
\def\thebibliography{\DeclareRobustCommand{\VAN}[3]{##3}\VANthebibliography}

\usepackage{graphicx}	
\usepackage{amsmath}	
\usepackage{caption}
\usepackage{subcaption}
\usepackage{lineno}

\newcommand{\mures}{$\mu_{\rm res}$}
\usepackage{eso-pic}

\AddToShipoutPictureBG*{%
  \AtPageUpperLeft{%
    \hspace{0.07\paperwidth}%
    \raisebox{-3.5\baselineskip}{%
      \makebox[0pt][l]{\textnormal{DES-2024-0829}}
}}}%

\AddToShipoutPictureBG*{%
  \AtPageUpperLeft{%
    \hspace{0.07\paperwidth}%
    \raisebox{-4.5\baselineskip}{%
      \makebox[0pt][l]{\textnormal{FERMILAB-PUB-24-0288-PPD}}
}}}%

\title[Dust in SNe Ia]{Modelling the impact of host galaxy dust on type Ia supernova distance measurements}

\author[B. Popovic]{
\parbox{\textwidth}{
\Large
B.~Popovic,$^{1}$\thanks{b.popovic@ip2i.in2p3.fr}
P.~Wiseman,$^{2}$
M.~Sullivan,$^{2}$
M.~Smith,$^{2}$
S. Gonz\'alez-Gait\'an,$^{3}$
D.~Scolnic,$^{1}$
J.~Duarte,$^{3}$
P.~Armstrong,$^{4}$
J.~Asorey,$^{5}$
D.~Brout,$^{7}$
D.~Carollo,$^{8}$
L.~Galbany,$^{9,10}$
K.~Glazebrook,$^{11}$
L.~Kelsey,$^{12,2}$
R.~Kessler,$^{13,14}$
C.~Lidman,$^{15,4}$
J.~Lee,$^{16}$
G.~F.~Lewis,$^{17}$
A.~M\"oller,$^{11}$
R.~C.~Nichol,$^{18}$
B.~O.~S\'anchez,$^{19,1}$
M.~Toy,$^{2}$
B.~E.~Tucker,$^{4}$
M.~Vincenzi,$^{1,2}$
T.~M.~C.~Abbott,$^{20}$
M.~Aguena,$^{21}$
F.~Andrade-Oliveira,$^{6}$
D.~Bacon,$^{12}$
D.~Brooks,$^{22}$
D.~L.~Burke,$^{23,24}$
A.~Carnero~Rosell,$^{25,21}$
J.~Carretero,$^{26}$
F.~J.~Castander,$^{9,10}$
L.~N.~da Costa,$^{21}$
M.~E.~S.~Pereira,$^{27}$
T.~M.~Davis,$^{28}$
S.~Desai,$^{29}$
S.~Everett,$^{30}$
I.~Ferrero,$^{31}$
B.~Flaugher,$^{32}$
J.~Garc\'ia-Bellido,$^{33}$
E.~Gaztanaga,$^{9,12,10}$
R.~A.~Gruendl,$^{34,35}$
G.~Gutierrez,$^{32}$
S.~R.~Hinton,$^{28}$
D.~L.~Hollowood,$^{36}$
K.~Honscheid,$^{37,38}$
D.~J.~James,$^{7}$
K.~Kuehn,$^{39,40}$
O.~Lahav,$^{22}$
S.~Lee,$^{30}$
J.~L.~Marshall,$^{41}$
J. Mena-Fern{\'a}ndez,$^{42}$
R.~Miquel,$^{43,26}$
J.~Myles,$^{44}$
R.~L.~C.~Ogando,$^{45}$
A.~Palmese,$^{46}$
A.~Pieres,$^{21,45}$
A.~A.~Plazas~Malag\'on,$^{23,24}$
E.~Sanchez,$^{47}$
D.~Sanchez Cid,$^{47}$
M.~Schubnell,$^{48}$
I.~Sevilla-Noarbe,$^{47}$
E.~Suchyta,$^{49}$
M.~E.~C.~Swanson,$^{34}$
G.~Tarle,$^{48}$
V.~Vikram,$^{6}$
and N.~Weaverdyck$^{50,51}$
\begin{center} (DES Collaboration) \end{center}
}
\vspace{0.4cm}
\\
\parbox{\textwidth}{
Affiliations shown at end of paper
}
}

\date{Accepted XXX. Received YYY; in original form ZZZ}

\pubyear{2024}

\begin{document}
\label{firstpage}
\pagerange{\pageref{firstpage}--\pageref{lastpage}}
\maketitle

\begin{abstract}
Type Ia Supernovae (SNe Ia) are a critical tool in measuring the accelerating expansion of the universe. Recent efforts to improve these standard candles have focused on incorporating the effects of dust on distance measurements with SNe Ia. In this paper, we use the state-of-the-art Dark Energy Survey 5 year sample to evaluate two different families of dust models: empirical extinction models derived from SNe Ia data, and physical attenuation models from the spectra of galaxies. Among the SNe Ia-derived models, we find that a logistic function of the total-to-selective extinction $R_V$ best recreates the correlations between supernova distance measurements and host galaxy properties, though an additional 0.02 magnitudes of grey scatter are needed to fully explain the scatter in SNIa brightness in all cases. These empirically-derived extinction distributions are highly incompatible with the physical attenuation models from galactic spectral measurements.
From these results, we conclude that SNe Ia must either preferentially select extreme ends of galactic dust distributions, or that the characterisation of dust along the SNe Ia line-of-sight is incompatible with that of galactic dust distributions. 
\end{abstract}

\begin{keywords}
cosmology:distance scale -- supernovae:general -- ISM: dust, extinction
\end{keywords}



\section{Introduction}\label{sec:Intro}

Type Ia supernovae (SNe Ia) have been critical tools in the measurement of the accelerating expansion of the universe \citep{Riess98, Perlmutter99}. This accelerating expansion may be driven by \lq dark energy\rq, parameterised by an equation-of-state $w$. Despite more than two decades of investigation, the nature of dark energy remains a cosmological mystery. 

To measure the accelerating
expansion, the brightness of SNe Ia must
be standardised in order to measure the distance to the SN \citep[e.g.,][]{Phillips93,Tripp98}. The largest standardisation correction accounts for the observation that redder SNe Ia are fainter and bluer SNe Ia are brighter (the \lq colour--luminosity\rq\ relation), and is based on measurements of the SN colours. This colour is likely to be a combination of an intrinsic SN colour and an extrinsic reddening due to dust along the line of sight to the SN. Some early SN Ia standardisation approaches attempted to separate the intrinsic colour and dust effects \citep{Riess96,Jha07}, assuming a phase-dependent intrinsic colour and an exponential distribution of dust reddening. Some modern methods also attempt the same separation \citep[e.g.][]{Mandel17,Burns11,Thorp21,Ward23}, but the commonly used SALT light curve model \citep{Guy05,Guy07} does not differentiate between the different astrophysical sources that affect the SN colour. Using the SALT standardisation framework to measure distances with SNe Ia assumes that intrinsic colour and extrinsic dust share the same colour--luminosity standardisation relationship (hereafter $\beta$).

\cite{BS20} suggest that SNe Ia may not be affected by a common $R_V$ -- the ratio of total to selective extinction caused by dust $A_V = R_V \times E_{\rm dust}$--  across all galaxy types and environments, and therefore the standardisation assumption of a universal $\beta$ may not be valid \citep{Gonzalez-Gaitan21}.  Instead, a variation in $R_V$ will cause a different amount of extinction for the same amount of reddening, resulting in a different effective value of $\beta$. If these $R_V$ differences are indeed host galaxy dependent, they may explain a number of otherwise puzzling observational effects in SN Ia data. These include i) the so-called \lq mass step\rq, the observation that SN Ia standardised brightnesses are 0.05-0.15\,mag fainter in low mass galaxies than in high-mass galaxies \citep{Kelly10,Sullivan10,Lampeitl10}, ii) the observation that $\beta$ decreases with increasing stellar mass \citep{Sullivan11}, and iii) the observation that the scatter in SN Ia Hubble residuals increases in SNe Ia with redder colours \citep[]{BS20}. 

Measuring the $R_V$ of any host-galaxy line-of-sight is difficult because of the degeneracy with the intrinsic spectral energy distribution (SED) of background sources. To mitigate the difficulty of directly inferring $R_V$ from SN Ia light curves, \citet{Popovic22} developed a Markov Chain Monte-Carlo technique to infer the $R_V$ of SN Ia populations and found a $\Delta R_V \sim -1$ between low and high mass galaxies is required to explain the Hubble residual versus SN colour trend. \citet{DES5YR} implement this technique on the Dark Energy Survey 5-year sample of SNe Ia and find a similar $\Delta R_V$ as well as a residual mass step of 0.04\, mag across all SN colours. \citet{Wiseman22} implemented a forward-modelling of the relationship between SNe Ia and their host galaxies, tracing SN Ia progenitors through a toy model of galaxy evolution through star formation history and stellar mass. They found similar $\Delta R_V$ values to \citet{Popovic22}, regardless of whether the $R_V$ varies as a function of stellar mass or mass-weighted galaxy age. 

These results, and the  \citet{BS20} model, find that a smaller $R_V$ is needed in high-mass galaxies \citep{Popovic21a,Popovic22}, or in galaxies with older stellar populations \citep{Wiseman22,Wiseman23}. In addition, the offsets between the $R_V$ of SNe Ia in low- and high-mass galaxies are large and assumed to be a step function. Independent observational evidence for such $R_V$ variation is scant, with most studies demonstrating any $R_V$ variation to be in the opposite sense. For example, \citet[][hereafter S18]{Salim18} measured dust attenuation in a large sample of star-forming and quiescent galaxies and found that amongst star-forming galaxies, the slope of the dust extinction law ($R_V$) {\it increases} as a function of stellar mass, the opposite sense to that inferred from the SN distance measurements. With specific star-formation rate (sSFR) there is a strong bimodality: low-sSFR, passive galaxies show $R_V\simeq 2.6$ whereas the mean for star-forming galaxies is 3.15, close to the average Milky Way value. This $\Delta R_V$ of $\sim0.5$ is significantly smaller than the $\sim1$ inferred from SN distance residuals by \citet{BS20,Popovic22,Wiseman22}. 

However, these $R_V$ values measured by S18 and within the Milky Way may not be directly comparable to the $R_V$ values derived from supernova measurements. Attenuation is not the same as extinction: it is the integrated effect of absorption and scattering both into and out of the line of sight to an unresolved ensemble of stars, whereas the extinction affecting an SN is purely a property of the line of sight to that SN \citep[see][for an investigation into the affects of attenuation on SN hosts]{Duarte22}.

Using the BayeSN lightcurve fitter, which leverages near-infrared (NIR) data to help constrain intrinsic colour and dust extinction in individual or ensembles of SN Ia light curves, \cite{Mandel22, Grayling24, Thorp21, Ward23} find conflicting evidence for a different $R_V$ between the populations of SNe in low- and high-mass galaxies. An additional complication is the question of whether $R_V$ variation can capture the full diversity of the observational trends. Recent works such as \citet{Rigault18}, \citet{Rose19}, \citet{Briday21}, \citet{Kelsey22} and \citet{Wiseman23}, have shown that properties related to the age of the stellar population local to the SN explosion site show the largest difference in SN Hubble residuals.

In this paper we attempt to reconcile the $\Delta R_V$ between SN Ia sight lines in low and high-mass galaxies inferred from the cosmological measurements to those measured in galaxy samples. We address the non-physical ``step" nature of the $R_V$ difference, and demonstrate that $R_V$ variations cannot account for the full intrinsic scatter of SN Ia distance moduli.

In Section \ref{sec:Data} we present the Dark Energy Survey and low-redshift supernova and host galaxy data used in this paper, followed by a review of dust models and the light-curve fitter in Section \ref{sec:Methods}. The dust models that we review are presented in Section \ref{sec:DustModels} and results are given in Section \ref{sec:Results}. Finally, the conclusions and discussion can be found in Section \ref{sec:Conclusions}.

\section{Data}\label{sec:Data}

In this paper we use the \lq 5-year\rq\ data release from the Dark Energy Survey (DES; \citealt{Flaugher15}) SN programme (DES-SN, Sanchez et al. \textit{in prep}). This release provides 1500 likely DES SNe Ia over $0.1<z<1.13$ with $griz$ light curves. Host galaxies are retrieved from deep co-added images \citep{Wiseman20}, and properties such as stellar mass and rest-frame colour are derived by fitting their spectral energy distributions with population synthesis templates \citep{Smith20, Kelsey22}. Each DES SN has a spectroscopic (host-galaxy) redshift from the Australian DES survey (OzDES) using the Anglo-Australian Telescope (Yuan et al., 2015, Childress et al., 2017, \citealt{Lidman20}), coupled with a photometric classification using the SuperNNova program \citep{Moller19}.

The DES-SN sample is complemented with external low redshift samples; CfA3 \citep{Hicken09b}, CfA4 \citep{Hicken12}, CSP \citep{Krisciunas17} (DR3) and the Foundation SN sample \citep{Foley17}. These samples comprise a range of $0.025<z<0.1$. Table~\ref{tab:data} shows a breakdown of the SNe Ia after quality cuts and light-curve fitting. A more thorough review of this selection is found in \cite{DES5YR, Moller24}. Of note, we do not include bias corrections on the SNIa distance modulus, as we are aiming to understand the underlying astrophysical relationships.

\begin{table}
\centering
\begin{tabular}{c|c}
\textbf{Cut} & \textbf{Total SNe} \\
\hline
SALT3 fit converged and $z>0.025$ &  3621 \\
$|x_1|<3$~\&~$|c|<0.3$ &  2687 \\
$\sigma_{x_1}<$1,~$\sigma_{t_{\rm peak}}<2$ &  2155 \\
\texttt{FITPROB}$>0.001$ &  2056 \\
Host Spec-$z$ & 1775 \\
$P_{\rm Ia} > 0.5$ & 1650 \\
\hline
\textbf{Final} & \textbf{1650} \vspace{1mm} \\
\end{tabular}
    \caption{SNe and Quality Cuts}
    \label{tab:data}
\end{table}

\begin{figure*}
    \centering
    \includegraphics[width=18cm]{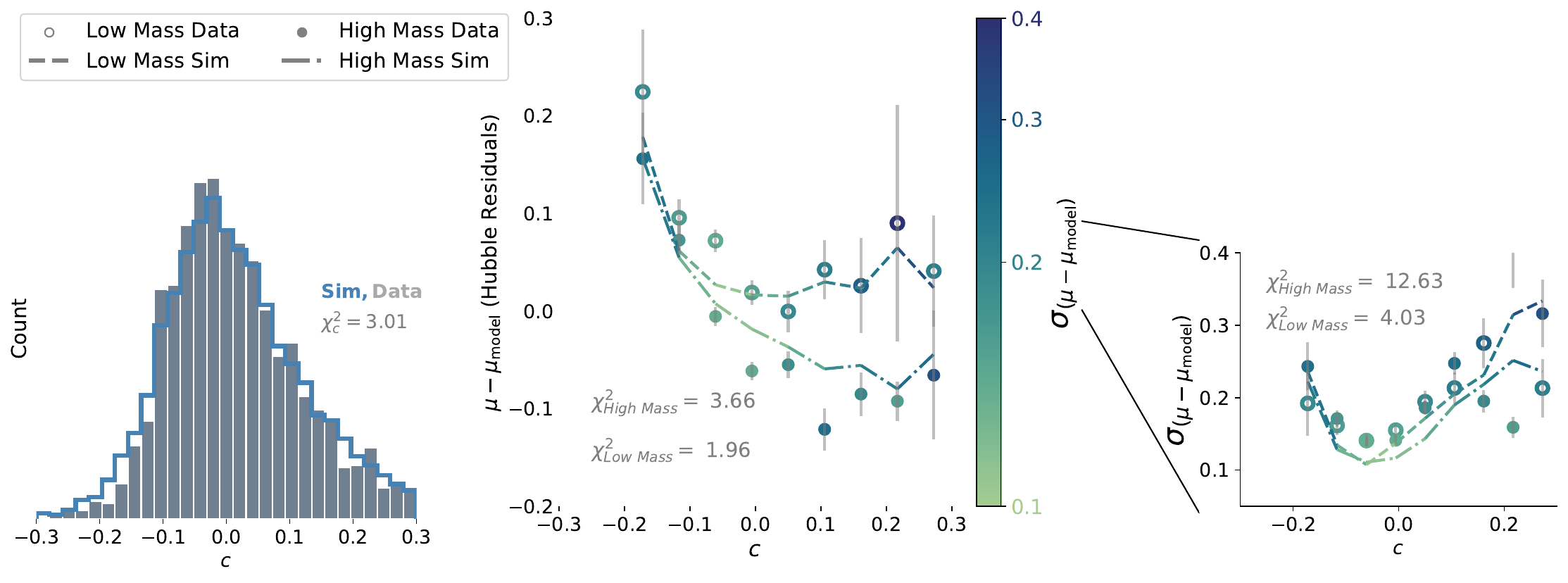}
    \caption{Plots of the metrics used to measure goodness-of-fit. Histogram of $c$, $\mu_{\rm res}$ as a function of $c$, and $\sigma_{\mu {\rm res}}$ as a function of $c$ are shown left to right. For the $c$ vs. $\mu_{\rm res}$ and $c$ vs. $\sigma_{\mu {\rm res}}$ plots, we split on the host galaxy mass $M_*$; high mass and low mass data are represented in closed and open circles respectively, and simulations are shown in dash-dotted and dashed line for high and low mass simulations. The $c$ vs $\mu_{\rm res}$ plot is colour coded by the Hubble scatter in each colour bin, this is elucidated in the rightmost figure. Here we present the DES5YR data and their nominal simulation.}
    \label{fig:data-sim}
\end{figure*}

\section{Analysis Methods}\label{sec:Methods}

The framework for the simulations presented in this paper has been developed primarily around the \textsc{snana} simulation software \citep{SNANA,Kessler19}, with the host galaxy forward model and parameter inference supplied by \citet{Wiseman22,Wiseman23} and \citet{Popovic21a,Popovic22} respectively. We briefly outline each of the procedures below.

\subsection{Simulations}\label{sec:Methods:subsec:Sims}

We use simulations to test our models; these simulations are generated with {\sc snana}. {\sc snana} broadly works in three steps: fluxes generated from a source model, addition of noise, and detection based on a characterisation of survey construction. 

Our base source model is the newest version of the Spectral Adaptive Light-curve Template (SALT3, \citealp{Kenworthy21}) model, an update from SALT2 \cite{Guy07}. SALT3 models the flux of an SN Ia as 
\begin{align} 
\label{saltmodel}
\begin{split}
F(\rm{SN}, p, \lambda) = x_{0} &\times\left[M_{0}(p, \lambda)+x_{1} M_{1}(p, \lambda)+\ldots\right] \\
&\times \exp [c C L(\lambda)],
\end{split}
\end{align}
where $x_0$ is the amplitude of the light curve, $x_1$ is the fitted light-curve stretch, and $c$ is the SN Ia colour parameter, similar to an $E(B-V)$ reddening. The $M_0$, $M_1$, and $CL(\lambda)$ parameters are determined for the trained model; each SN Ia has a fitted $x_0$, $c$, and $x_1$.

Distances are inferred from SALT3 via the Tripp estimator \citep{Tripp98}. The distance modulus is given as 
\begin{equation}
\label{eq:tripp}
    \mu = m_B + \alpha_{\rm SALT} x_1 - \beta_{\rm SALT} c - M_0
\end{equation}
where $m_B = -2.5\textrm{log}_{10}(x_0)$ and $c$ and $x_1$ are defined above. The $\alpha_{\rm SALT}$ and $\beta_{\rm SALT}$ are sample-dependent nuisance parameters, following \cite{Kenworthy21}. $M_0$ is the absolute magnitude in the $B$-band of an SN Ia with $c=x_1=0$. We fit for $\alpha_{\rm SALT}$ and $\beta_{\rm SALT}$ when testing each of our models.

\subsection{Review of treatment of dust in simulations of SN Ia populations }\label{sec:Methods:subsec:Dust}

The aim of this paper is to test the efficacy of dust models; therefore here we will briefly review the dust model methodology introduced in \cite{BS20} and updated in \cite{Popovic22}.


Dust models for SNe Ia attribute the distribution of SN Ia colours to an intrinsic, dust-free colour component $c_{\rm int}$ that is reddened by a dust component. The observed SN Ia colour ($c$ in Eq. \ref{eq:tripp}) is then modeled as 
\begin{equation}
    c = c_{\rm int} + E_{\rm dust} + \epsilon_{\rm noise}
\label{eq:cobs}
\end{equation}
where $E_{\rm dust}$ is the dust component and where $\epsilon_{\rm noise}$ is otherwise unaccounted-for measurement noise. 

The component $E_{\rm dust}$ from Eq.~\ref{eq:cobs} is interpreted as $E(B-V)$, such that the $V-$band extinction is given by 
\begin{equation}
    A_V = R_V \times E_{\rm dust}
\label{eq:avdustmodel}
\end{equation}
where $R_V$ is total-to-selective extinction ratio and $E_{\rm dust} = E(B-V)$. The change in observed brightness due to colour, i.e. what is fit as $\beta_{\mathrm{SALT}}c$ (from Eq. \ref{eq:tripp} can be decomposed into 
\begin{equation}
    \Delta m_B = \beta_{\rm SN}c_{\rm int} + (R_V+1)E_{\rm dust} + \epsilon_{\rm noise}.
    \label{eq:deltamb}
\end{equation}
where $\beta_{\rm SN}$ is the intrinsic colour-luminosity relationship, and extinction acts as $R_B = R_V + 1$ in the $B$-band.

Further review can be found in \cite{BS20} and \cite{Popovic22}.

\subsection{Host Galaxies}\label{sec:Methods:subsec:Hosts}

Host galaxies are simulated using the physically-motivated empirical model of \citet{Wiseman22} using the updated prescription of \citet{Wiseman23}. A  full description of the simulations can be found in those works. Briefly, seed galaxies evolve following empirical relations that govern their build up of stellar mass (i.e. star-formation history) following the method of \citet{Childress14}. SNe are associated with galaxies following a probability distribution governed by realistic rates of SNe, themselves driven by the convolution of the SFH of each galaxy and the delay-time distribution of SNe. SNe are designated ``young” (from stellar populations with ages $<1$\,Gyr) or ``old” ($t>1$\,Gyr). The relative number of young and old SNe matches observations well, assuming that SN stretch is driven by this age distribution \citep{Rigault2020,Wiseman22}.

\section{Host galaxy dust relationships}\label{sec:DustModels}

Here we outline the models that we test in this paper, which are broadly split into two families. With the exception of the S18 model, our models of dust extinction are derived from empirical measurements of SNe Ia lightcurves, and are likely tracers of the line-of-sight \textit{extinction}; this is in contrast to S18, which provides measurements of the \textit{attenuation} of light due to dust.

In each of the following sections we lay out a different approach for testing distributions of $R_V$ and $E(B-V)$ and how they relate to host galaxy properties. The four remaining `dust-free' parameters that characterise the intrinsic properties of the SN Ia scatter model are the mean and standard deviation for the Gaussian distribution of $c_{\rm int}$ and $\beta_{\rm SN}$. Here we make note that $\beta_{\rm SN}$ is not the same as the one from the $\beta_{\rm SALT}$ in Eq. \ref{eq:tripp}: $\beta_{\rm SALT}$ from Eq. \ref{eq:tripp} is fit from the data, and is a convolution of $\beta_{\rm SN}$ and other dust effects.

We fix the intrinsic colour and intrinsic $\beta_{\rm SN}$ by using a single population that follows a Gaussian distribution for each; we use the values from \cite{DES5YR}: $\mu_c = -0.7$, $\sigma_c = 0.053$; $\mu_{\beta} = 2.07$, $\sigma_{\beta} = 0.22$. The only exception is for the WH23 model, which is further described later. If SNe actually belong to two intrinsic colour distributions, this could explain some or all of the mass step instead of different $R_V$ or $\beta$ populations. However, the DES5YR data are entirely consistent with a single colour population \citep{DES5YR}. We thus focus here on dust properties and defer the testing of multiple colour populations to future analyses.

\subsection{Baseline model: DES}

Our baseline simulation model uses the model parameters from the DES 5-year cosmological results \citep{DES5YR}. These parameters, which describe distributions for $c_{\rm int}$, $R_V$, $E_{\rm dust}$, and $\beta_{\rm SN}$, are simultaneously fit using the \texttt{Dust2Dust} program from \cite{Popovic22}, providing two populations of $R_V$ ($\mu_{\rm RV_{\rm high}} = 1.66$, $\mu_{\rm RV+{\rm low}} = 3.25$) that are split on the host galaxy stellar mass, specifically at log$(M_{*}) = 10$. The data used for the \texttt{Dust2Dust} training process have a cut on the photometric classification probability that the light curve is a SN Ia, $P_{\rm Ia} > 0.5$, applied, but the resulting parameters in \cite{DES5YR} are compared to the full cosmological data set in that work. Here, we have re-instituted the $P_{\rm Ia} > 0.5$ cut, as we wish to similarly avoid non-Ia contamination in the testing of our model parameters.

\subsection{Linear $R_V$ and $E(B-V)$ (P24c)}\label{sec:DustModels:subsec:ZTF}

Popovic et al. ({\it in prep}, hereafter P24c) compiles a volume-limited sample of SNe Ia from DES, the Zwicky Transient Facility, the Sloan Digital Sky Survey, and Pan-STARRS photometric SN Ia samples. To this collected sample, they fit a Gaussian intrinsic colour and an exponential $E(B-V)$ dust tail, described by the parameter $\tau$, to their SN Ia colour distribution as a function of host galaxy mass and redshift. This provides a statistical probability of the mean reddening, $\tau$, given a redshift and host galaxy mass for each supernova. Here, we use their exponential reddening values ($\tau$).

Furthermore, P24c similarly splits their data into uniform bins of host galaxy stellar mass, starting at log$(M_{*}) = 8$ to log$(M_{*}) = 12$, in steps of log$(M_{*}) = 0.5$. In each of these bins, they fit a $\beta_{\rm SALT2}$ value to the data, providing a measurement of a mass-varying $\beta_{\rm SALT2}$. We use these $\beta_{\rm SALT2}$ fits to describe the relationship of $R_V$ to host galaxy mass, assuming $\beta_{\rm SALT} = R_V + 1$ to correspond to the $B$-band. 

The P24c model, therefore, uses a linear $\beta_{\rm int}$ distribution and a finer-binned $\tau$ distribution, compared to the two Gaussian $R_V$ and two exponential $\tau$ distributions of the fiducial DES model. 

\subsection{Logistic $R_V$ curve (Logistic)}\label{sec:DustModels:subsec:Logistic}
It is unphysical to assume $R_V$ to be governed by a simple step function, whether with stellar mass or any other continuous host galaxy property. Here we model the $R_V$ distribution of our sample as a function of host galaxy mass $M_*$ using a logistic function:
\begin{equation}
    R_V = \frac{1.5}{1+e^{2({\rm log}(M_{*})-10)}} + 2
    \label{eq:logistic}
\end{equation}
where $L = 1.5$ and $k=2$.

\begin{figure}
\includegraphics[width=8.75cm]{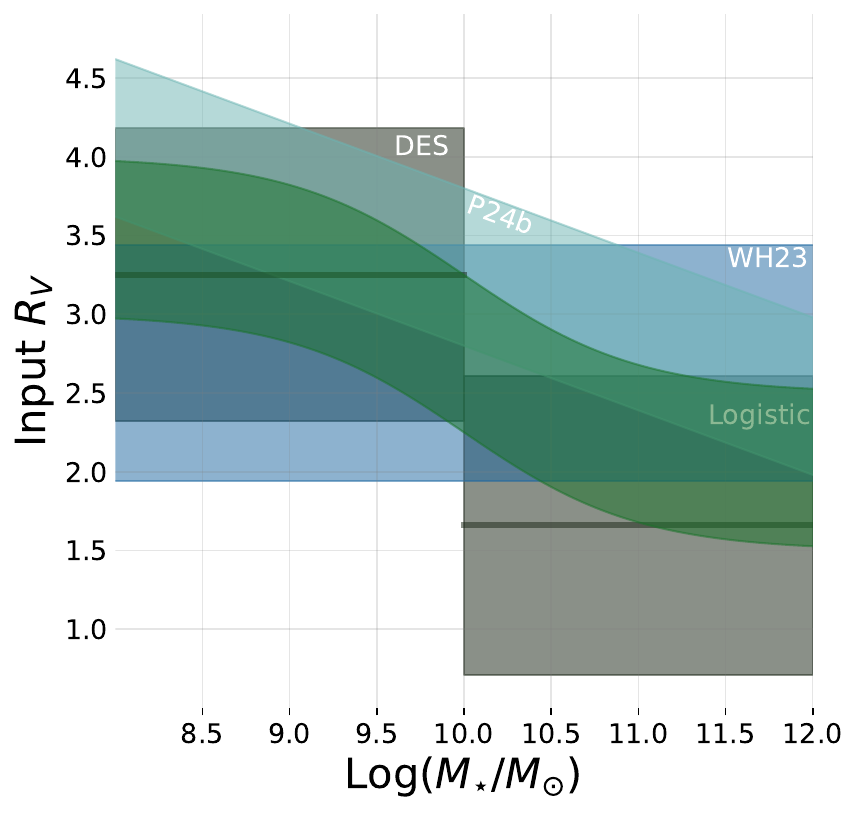}
\caption{An illustration of the input $R_V$ values for our SNIa inferred models. The fiducial DES model in dark grey assumes two Gaussian distributions split on log($M_*/M_{\odot}$)$=10$, as opposed to the continuously-varying P24c (turqoise) and Logistic (dark green) models. The light blue WH23 model does not vary with the host galaxy stellar mass.}
\label{fig:EXAMPLERV}
\end{figure}

We choose the logistic curve as a smoothly varying $R_V$ that spans the range of $R_V =$ 3.5 in the extreme low host galaxy mass (log$(M_*) < 8$) to 2.0 in the extreme high host galaxy mass (log$(M_*) > 12$). We add a Gaussian error with $\sigma = 0.5$ to the $R_V$ values. Our choice of an $R_V$ threshold of 2 is motivated to avoid the $R_V = 1.2$ Rayleigh scattering threshold, while maintaining a smooth function. The $L = 1.5$ and $k = 2$ values were chosen to mimic the $\Delta R_V = 1.5$ range and transition from other dust models, and are shown in Figure \ref{fig:EXAMPLERV}. We use the same $\tau$ values from P24c to describe the exponential reddening for this and subsequent tests.

\subsection{Salim et al. 2018 (S18)}\label{sec:DustModels:subsec:Samir}

We obtain the specific Star Formation Rate / Host Galaxy Mass / $R_V$ contours from \cite{Salim18} (S18) and include the $R_V$ information in our host library. S18 performs an SED fit on 230,000 galaxies using photometry from \textit{GALEX}, \textit{SDSS}, and \textit{WISE}. This SED fitting across multiple bands allows them to constrain star formation and $R_V$ across a range of galaxies from quiescent to star forming systems. We specifically use the data from the "Slope (all galaxies)" contour in Figure 3 in S18. We use the $E(B-V)$ distribution from P24c for this test.

\subsection{WH23}

\citealp{WH23} (WH23) performs a novel Bayesian Hierarchical model of SNe Ia parameters from the \textit{SuperCal} sample \citep{Scolnic15}. In contrasts to other works, they find evidence for \textit{no} variation in dust populations between host galaxies, instead finding two distinct intrinsic populations of $x_1$ and $c$. Additionally, even for a single-population model, WH23 infers a consistent Gaussian $R_V$ population centered at $\overline{R_V} = 2.96$ across host galaxy properties, and a novel two-tailed $E(B-V)$ distribution:
\begin{equation}
    p(E(B-V)) = x^{\gamma - 1} \exp{-x/\tau}
\end{equation}
where $\gamma = 3.13$ and $\tau = 0.035$. Here we test their `single population' model, with model parameters from Table 2 of WH23, with an eye towards the efficacy of the gamma distribution. Of note, their model requires a change in the intrinsic colour and $\beta$ values: we reproduce those changes in this paper by setting $\overline{c}_{\rm int} = -0.1$ and $\beta = 3.19$.

\subsection{Logistic and S18 with Intrinsic Step (+Step)}\label{sec:DustModels:subsec:SNSTEP}

Here we repeat the previous two models in Sections \ref{sec:DustModels:subsec:Logistic} (Logistic) and \ref{sec:DustModels:subsec:Samir} (S18), but with the addition of a luminosity step as a function of SN Ia age. We place our "age step" at log$_{10}({\rm SN~age}/1\,{\rm Gyr}) = 1$, following \citet{Wiseman22} and Wiseman et al. {\it in prep}. 

Additionally, we test the effects of an increasing age step on our Logistic $R_V$ model, with particular focus on how this luminosity step affects our Hubble Residual scatter. We increase the age step in steps of $0.08$ magnitudes, from $0$ to $0.32$ mags.

Works such as \cite{Rigault18,Kelsey22, Briday21,Wiseman22,Wiseman23} suggest that a luminosity step may be driven by properties other than the host galaxy stellar mass; we adopt this assumption in order to test not only the recovery of the mass step, but also the hypothesis that there may be a luminosity step that is driven by processes for which the host galaxy stellar mass acts as a biased tracer. Other works, such as \cite{Gonzalez-Gaitan21}, suggest that the luminosity step may arise from two separate populations of intrinsic colour. This model is incompatible with the baseline simulations from DES and would require a simultaneous fit of the two intrinsic populations and the dust distributions, which will be left to another paper. 

\subsection{$R_V$ Variation}\label{sec:DustModels:subsec:Rv}

Here we investigate the impact of increasing scatter in the $R_V$ distribution on our dust models. We again use the Logistic $R_V$ curve as a starting point and add a Gaussian scatter to the $R_V$ values in our host galaxy library. The first test begins with 0 scatter in the $R_V$ distribution, and we increase the $\sigma_{R_V}$ in each test by steps of $0.2$, up to a maximum value of $0.8$. 

\section{Results}\label{sec:Results}

To determine the efficacy of our models, we follow the criteria and method provided by \cite{Popovic22}. While detailed further in that paper, we briefly detail the three criteria here, whereby the $\chi^2$ are computed between simulations and data: 

\begin{itemize}
    \item $\chi^2_c$: Comparison of the SN Ia $c$ distribution, with Poisson errors.
    \item $\chi^2_{\mu {\rm res}}$: The $c$ vs. Hubble Residual $\mu_{\rm res}$ ($\mu_{\rm res} = \mu - \mu_{\rm model}$) curves, split on high and low mass (split at $10 M_*$), with $e_{\mu_{\rm res}} = \sigma / \sqrt{N}$, where $\sigma$ is the standard deviation of the Hubble Residuals.
    \item $\chi^2_{{\rm RMS}}$: The scatter in Hubble Residuals as a function of $c$, split on high and low mass (split at $10 M_*$). We measure the scatter with the median absolute deviation, and $e_{\rm RMS} = \sigma / \sqrt{2N}$.
\end{itemize}

Of note, we use a finer colour binning than \cite{Popovic22} and report the reduced $\chi^2$, $\chi^2_{\nu}$ for $\nu = 10$ degrees of freedom, for the ten uniform colour bins. The baseline model that we use, from \cite{DES5YR}, is presented in Figure \ref{fig:data-sim} alongside our three metrics.

The $\chi^2_{\nu}$ values for all of our tested models are given in Table \ref{tab:results}, and we go over in further depth here. From here, we report our reduced $\chi^2/\nu$ values without the $\nu$ denominator for visual clarity.

\subsection{Baseline}\label{sec:Results:subsec:DES}

Despite some changes in the data between this analysis and \cite{DES5YR}, notably $P_{Ia} > 0.5$ and $z < 0.7$ cuts, we find good agreement between our baseline model from DES and the data. 

\subsection{Popovic et al. 2024b (P24c)}\label{sec:Results:subsec:ZTF}

We find relatively good agreement between the P24c simulations and data. While P24c does contain DES data, the DES data only accounts for $\sim 1/3$ of the P24c training set. The overall performance of P24c is comparable to the baseline DES, and neither models the high mass Hubble Residual scatter well.

\subsection{Logistic $R_V$ Curve (Logistic)}\label{sec:Results:subsec:Logistic}

We find that the Logistic $R_V$ performs well in replicating the data. The logistic curve improves on both the \mures and Hubble Residual scatter curves.
Logistic $R_V$ does not replicate $\mu_{\rm res}$ vs. $c$ as well as the DES5YR baseline in the red $c>0$ regime but performs better in the blue $c<0$ where it reproduces roughly half of the mass step.

\subsection{Salim et al. 2018 (S18)}\label{sec:Results:subsec:Samir}

Overall, S18 performs comparably to our base model. However, S18 performs worse in modeling \mures. Current dust models attribute the mass step to differences in the mean of the $R_V$ distribution when split on mass: S18 does not produce a large difference: $\Delta R_V = \sim 0.5$. Therefore, S18 does not well reproduce the mass step, unlike other models.

\subsection{WH23}

The WH23 does not well-match the data. For the moment, we set aside the discussion of the $\chi^2_{\mu res}$, as we do not use the two-population model. The choice of a two-tailed dust distribution fails to replicate the observed colour distribution, even with the adjustment to the intrinsic colour distribution: $\overline{c}_{\rm int} = -0.1$. The increased average dust results in a significantly worse ($\times 2$) match to the high mass $\chi^2_{\rm RMS}$ criteria.

\subsection{Logistic and S18 with Intrinsic Step (+Step)}\label{sec:Results:subsec:SNSTEP}

We find that the addition of an intrinsic luminosity step that is dependent on the age of the SN Ia presents a small but overall improvement to our models. Table \ref{tab:stepSize} shows the Hubble Residual scatter $\chi^2_{\rm RMS}$ and total $\chi^2$ for our tested magnitudes of the age step. In our coarse search, we find that a step size $\gamma = 0.16$ returns the best $\chi^2$ value, and that the majority of this improvement in modeling the $\chi^2$ comes from improved matches in the Hubble Residual scatter.
We emphasise that we have modelled this additional step as an intrinsic difference in $M_0$ but it could equally well come from two intrinsic colour populations, or a combination of both. The source of this step is not addressed further in this work.

\subsection{$R_V$ Variation}\label{sec:Results:subsec:RV}

We test the impact of increased scatter of $R_V$ values on our data, by adding increasing amounts of Gaussian scatter onto the Logistical $R_V$ function. Table \ref{tab:sigRv} shows the impact of the $\sigma R_V$ choice on our logistic $R_V$ model. There is not a strong correlation between $\sigma R_V$ and the total $\chi^2$; higher $\sigma R_V$ such as $\sigma R_V \geq 0.8 $, are ruled out by the increased $\chi^2$. In contrast to the $\gamma$, the Hubble Residual scatter $\chi^2_{\rm RMS}$ is largely static below $\sigma R_V < 0.5$. We find that our best-fit $\sigma R_V = 0.5$, though again stress that this is still a coarse fit. 

\begin{table*}
    \centering
    \begin{tabular}{c|c|c|c|c|c|c|c}
         \textbf{Model} & \cite{DES5YR} (DES, base)  & P24c & Logistic $R_V$(Logistic) & S18 & Logistic+Step & S18+Step & WH23  \\
         \hline
         $\chi^2_{c}$                     & 3.0  & 2.9 & 2.5 & 2.5 & 2.5 & 2.5 & 13.2  \\
         High Mass $\chi^2_{\mu \rm res}$ & 3.7  & 4.2 & 3.3 & 6.4 & 2.7 & 5.7 & 9.43 \\
         Low Mass $\chi^2_{\mu \rm res}$  & 2.0  & 3.7 & 1.5 & 8.6 & 1.0 & 6.3 & 8.51 \\
         High Mass $\chi^2_{\rm RMS}$     & 12.6 & 14 & 12 & 13 & 11 & 12 & 27.62 \\
         Low Mass $\chi^2_{\rm RMS}$      & 4.0  & 1.0 & 3.0  & 5.0 & 3.0 & 5.0 & 3.84 \\
         \textbf{Total $\chi^2$} & \textbf{25.3} & \textbf{25.8} & \textbf{22.3} & \textbf{35.5} & \textbf{20.2} & \textbf{31.5} & \textbf{62.5} \\
         
    \end{tabular}%
    \caption{$\chi^2/\nu$ values for the dust models presented in this work, where $\nu = 27$.}
    \label{tab:results}
\end{table*}

\begin{table}
    \centering
    \begin{tabular}{l|c|c}
        Model & $\chi^2_{\rm RMS}$ & \textbf{Total $\chi^2$} \\
        \hline
        $\sigma R_V = 0.5$ (reference) & 15.3 & \textbf{22.7} \\
        $\sigma R_V = 0$ & 19.6 & \textbf{27.1} \\
        $\sigma R_V = 0.2$ & 19.2 & \textbf{27.0} \\
        $\sigma R_V = 0.4$ & 19.8 & \textbf{27.3}\\
        $\sigma R_V = 0.6$ & 16.7 & \textbf{25}\\
        $\sigma R_V = 0.8$ & 22.3 & \textbf{30.7}\\
    \end{tabular}
    \caption{$\chi^2$ results for increasing values of $\sigma R_V$, tested with the Logistic $R_V$ distribution.}
    \label{tab:sigRv}
\end{table}

\subsection{Recovery of $\beta_{\rm SALT2}$}\label{sec:Results:subsec:Beta}

\begin{figure}
\includegraphics[width=8.75cm]{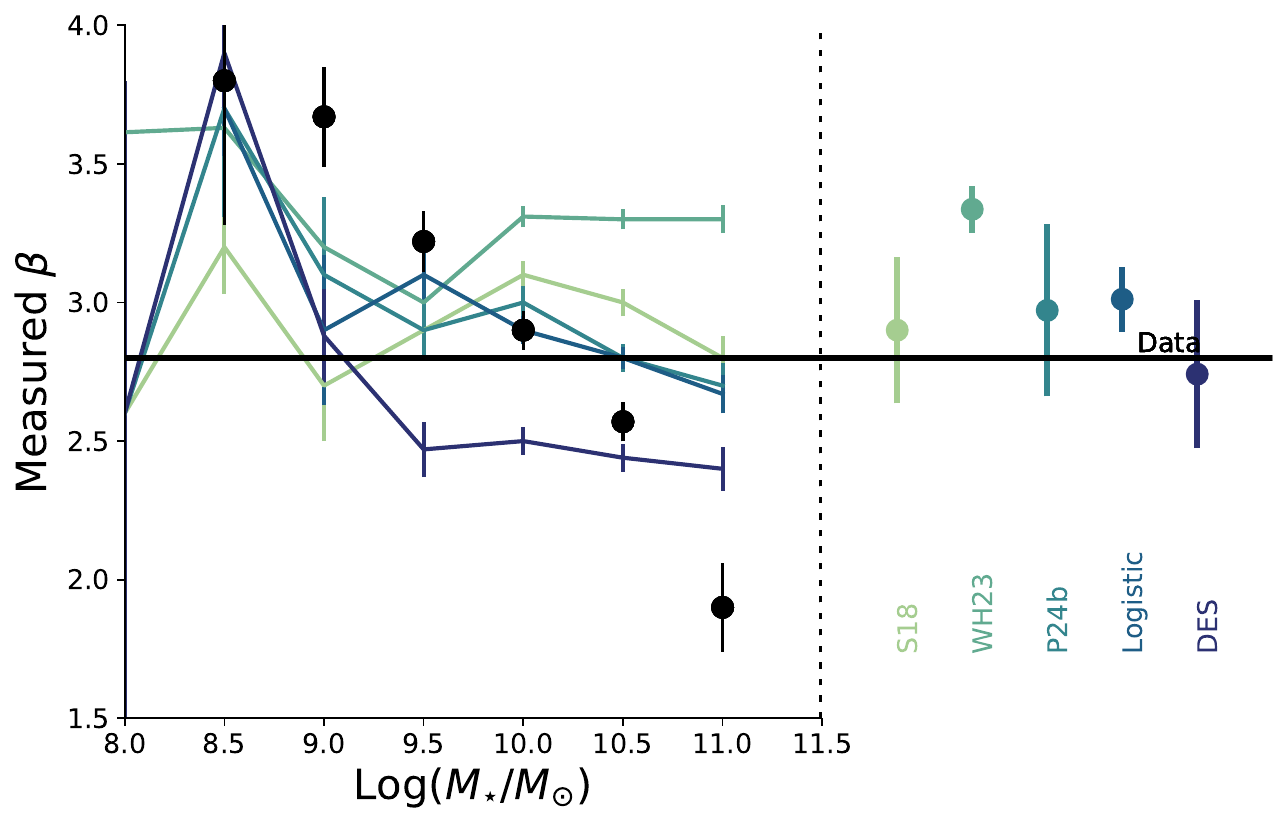}
\caption{The $\beta$ vs host galaxy mass relationship for the DES data (black points) and each of our tested models (coloured lines).  On the right, the conventional Tripp $\beta$, which is marginalised over mass, is presented for each model.}
\label{fig:BVM}
\end{figure}

Figure \ref{fig:BVM} shows that the DES data display the same relationship between $\beta_{\rm SALT}$ and host stellar mass as P24c: beta decreases with increasing host stellar mass.
The recovered Tripp $\beta$ for each of our models is well-matched to the data. In contrast, none of our models are able to entirely recover the range of the observed $\beta$ vs. host galaxy mass relationship seen in the data, though each model is well-able to replicate the $\beta$ value when marginalising over the host galaxy stellar mass. 

\begin{table}
    \centering
    \begin{tabular}{l|c|c}
        Model & $\chi^2_{\rm RMS}$ & \textbf{Total $\chi^2$} \\
        \hline
        $\gamma  = 0$ (reference) & 15.3 & \textbf{22.7} \\
        $\gamma  = 0.08$ & 12.6 & \textbf{18.9} \\
        $\gamma  = 0.16$ & 10.5 & \textbf{16.4} \\
        $\gamma  = 0.24$ & 25.4 & \textbf{33.6} \\
        $\gamma  = 0.32$ & 76.8 & \textbf{88.8} \\
    \end{tabular}
    \caption{$\chi^2$ results for increasing values of $\gamma$, tested with the Logistic $R_V$ distribution.}
    \label{tab:stepSize}
\end{table}

\section{Discussion and Conclusions}\label{sec:Conclusions}

In this paper we have reviewed a number of different methods to model dust distributions and properties for use in standardisation of SNe Ia, and evaluated their efficacy. We have included an age-dependent luminosity step that is independent of the dust of the host galaxy to test alongside these methods, and compared them to dust-only methods. Here we will discuss our results.

\subsection{SN Ia Hubble Residual Scatter}\label{sec:Conclusions:subsec:Scatter}

The single largest $\chi^2$ contribution across all models we have tested is the high mass Hubble Residual scatter, $\sigma_{\mu {\rm RMS}}$. We do see slight improvements to the high mass $\sigma_{\mu {\rm RMS}}$ $\chi^2/\nu$ values when including an age-based luminosity step, up to a $\chi^2/\nu$ reduction of $\sim 2$ in the case of $\gamma = 0.16$. Higher values are ruled out by an increasing $\chi^2_{\rm RMS}$, putting an upper limit on this value.

The $\sigma R_V$, variation, on the other hand, does not present a strong constraint on our models. Firstly, this would indicate that variations in $R_V$ are not responsible for the observed Hubble residual scatter, as we would expect to see improved fits to the Hubble residual scatter with a greater $R_V$ variation. Instead, we find that any $\sigma R_V \leq 0.4$ is equally supported by the data; likely a sign of a detection threshold in the data, that we cannot detect a variation in $\sigma R_V$ below $0.4$.

Overall, all of our tested models noticeably underestimate the scatter seen in SNe Ia. The addition of a colour/mass-independent scatter floor of $\sim 0.02$ improves our $\sigma_{\mu {\rm RMS}}$ $\chi^2$ values by a factor of approximately two. This can be compensated by an age step, indicating that dust is not the only element at play in the observed SN Ia scatter. 

\subsection{SN Ia Dust Distributions}\label{sec:Conclusions:subsec:Dust}

The logistic $R_V$ distribution provides an improved model for dust to be the sole contributor to SN Ia scatter and the mass step. Nonetheless, the model is unable to replicate the necessary level of Hubble Residual scatter, and unable to recreate the mass step. 

Approaching from the alternative direction of galaxy observations, we find that SNe Ia with host galaxy dust properties drawn from the global galaxy population do not well-match either of our Hubble Residual or scatter metrics, failing to recreate the mass step. This is because after mapping the SN Ia host galaxy stellar mass function onto the S18 galaxy $R_V$ values, there is an \textit{increase} in $R_V$ with the host galaxy stellar mass, contrary to the predictions of contemporary dust models (including our Logistic function), which require \textit{lower} $R_V$ in more massive hosts. 

\begin{figure}
    \centering
    \includegraphics[width=8.75cm]{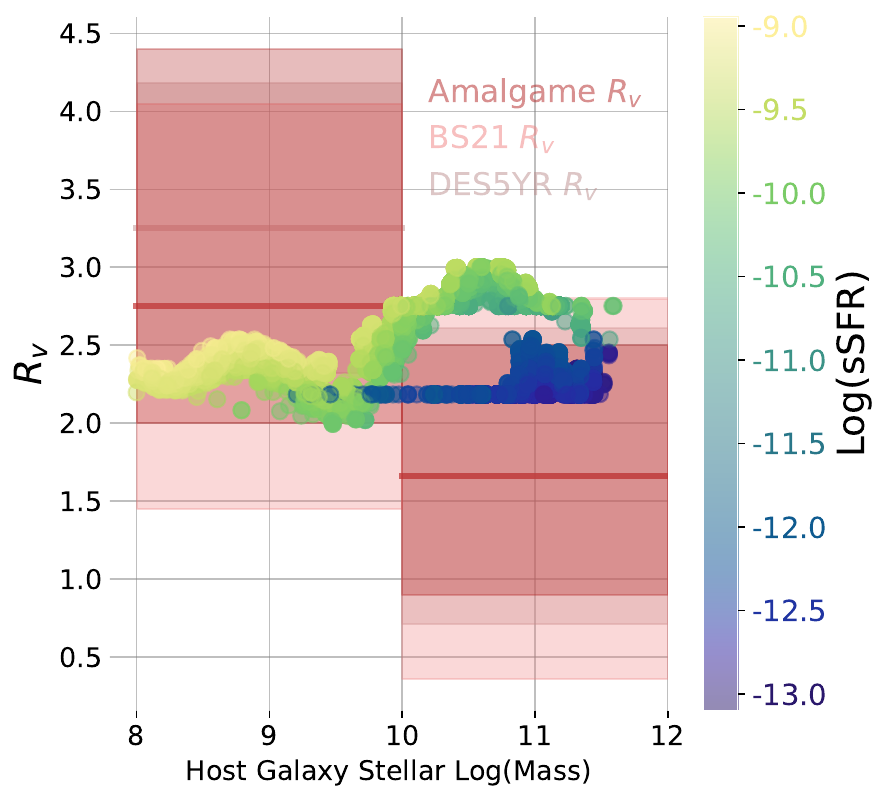}
    \caption{The simulated $R_V$ versus host galaxy stellar mass relationship colour coded by Log(sSFR), based on the galaxy data from S18. The trend of increasing $R_V$ with host galaxy stellar mass is opposite to that predicted by contemporary SN Ia dust-based scatter models, which are shown in shades of red. The high-mass $R_V$ values for the contemporary SN Ia dust models overlap strongly.}
    \label{fig:S18Rvs}
\end{figure}

The S18 results indicate that sSFR, rather than stellar mass, does show a change of $R_V$ in the sense required by the SN Ia distances: a galaxy at fixed stellar mass will have a lower $R_V$ if it has a lower sSFR. Nevertheless, this trend works against the trend of increasing $R_V$ with stellar mass in the star-forming galaxies, such that the low mass star-forming galaxies and high mass passive galaxies have comparable $R_V$ values. We illustrate this in Fig.~\ref{fig:S18Rvs}, which shows the $R_V$ verus stellar mass relationship for simulated SNe Ia, colour coded by the $\log({\rm sSFR})$ value of the galaxy. Even excluding all high-mass star-forming galaxies, the largest $\Delta R_V$ between low and high mass galaxies is $\sim 0.3$, less than the values needed to recreate the mass step with $R_V$ alone: $\Delta R_V = \sim 1$. The presence of star-forming galaxies will serve only to increase the median $R_V$ value in high-mass galaxies. 

\subsection{Dark dust}
Instead of an intrinsic luminosity difference causing the unexplained step in SNe of all colours, an offset could be introduced by so-called `dark dust'. Dark dust (e.g. Siebenmorgen 2023) is not selective in its extinction, such that all wavelengths are equally extinguished. If there were systematically more dark dust along sight lines to SNe Ia in young star-forming environments than those in older passive environments, then those SNe would be systematically fainter regardless of their colour. There is tentative evidence for such a trend based on the emission from cold dust grains in passive galaxies (Krügel et al. 1998; Siebenmorgen et al. 1999).

We conclude that if a varying $R_V$ is responsible for the SN Ia mass step and its evolution with SN colour, then the extinction along SN Ia lines of sight, and its relationship with the host galaxy properties, is not well reproduced by the corresponding global attenuation of SN Ia host galaxies.
There are four possible explanations for this behaviour:
\begin{itemize}
    \item the attenuation measured from integrated galaxy observations, and how it relates to global galaxy properties, is not representative of how line-of-sight extinction varies with these properties, and/or
    \item SNe Ia are preferentially observed along lines of sight at the extreme ends of dust distributions, and/or
    \item Galaxies contain significant quantities of dark dust, and there is systematically more in low-mass, star-forming environments, and/or
    \item There is an intrinsic brightness difference in SNe Ia that evolves with SN colour but is not caused by dust, such as two populations of $M_0$ or $c_{\rm int}$.
\end{itemize}
The inability of our dust-only models from the literature to recreate the mass step appears to match with near-infrared measurements of SNe Ia, particularly \cite{Thorp21} and \cite{Ward23}, both analyses of SNe Ia with the BayeSN light curve fitter. These studies do not find a significant $\Delta R_V$ between high and low mass galaxies. 

We conclude that a portion of the observed mass step is likely caused by an intrinsic brightness difference closely related to the age of the SN progenitor and its local environment, and that an especially large $\Delta R_V$ $( > 1)$ is not supported by galactic $R_V$ distributions. This may be due to SNe Ia preferentially selecting the extremes of an $R_V$ distribution, but it is more likely that the method of measuring $R_V$ and $E(B-V)$ for SNe and galaxies are incompatible. Measuring the dust of a supernova involves a single line of sight and can be described as extinction, whereas a non-point source measurement across a galaxy requires accounting for attenuation and scattering; this is described further in \cite{Chevallard13,Narayanan18,Duarte22}.

\subsection{Conclusions}\label{sec:Conclusions:subsec:Conclusions}

The discrepancy between the SN Ia extinction and galactic attenuation presents a new difficulty for using the host galaxy properties of an SN Ia to standardise its distance, and further work must be done to determine a way to connect the host galaxy dust properties to the dust that is along the line of sight to the supernova. It appears that initial attempts to directly provide a one-to-one correlation between the properties of the host galaxy and its supernova will need continuing research.

The introduction of an age step partially ameliorates the need of an otherwise quite large $\Delta R_V$ between high and low mass galaxies, but this benefit primarily occurs in our modeling of the Hubble Residual scatter, rather than the mass step itself. That the size of the age step does not make a large impact on the mass step is explained by Wiseman et al., {\it in prep}, but points to further work on disentangling different tracers of anomalous luminosity offsets.

In the shorter term, incorporating dust into SN Ia standardisation has provided benefits beyond modeling the mass step, and there are clear benefits to their continued use \citep{BS20, Brout22, Popovic21a, Popovic24a}. We provide an improvement to the $R_V$ step used in previous literature (e.g. \cite{BS20, Wiseman22, Popovic22}) via the use of a smoothly varying logistic function that performs better than a linearly-varying $R_V$ distribution. Here we have not rigorously optimised the logistic $R_V$ nor the age step size; such work ought to be incorporated into pipelines such as UNITY \citep{UNITY} or Dust2Dust \citep{Popovic22}.

\section*{Acknowledgements}
All authors have read and contributed to the drafting of the manuscript. BP devised the analysis, ran the simulations and drafted the majority of the manuscript. PW, MSu and MSm provided scientific support throughout the analysis and writing. SG-G and DS internally reviewed the work and provided extensive feedback. JD, LG, LK, RK, CL, JL, MT and MV provided comments on the analysis and interpretation. The remaining authors have made contributions to this paper that include, but are not limited to, the construction of DECam and other aspects of collecting the data; data processing and calibration; developing broadly used methods, codes, and simulations; running the pipelines and validation tests; and promoting the science analysis.

PW acknowledges support from the Science and Technology Facilities Council (STFC) grant ST/R000506/1. AM is supported by the ARC Discovery Early Career Researcher Award (DECRA) project number DE230100055

Funding for the DES Projects has been provided by the U.S. Department of Energy, the U.S. National Science Foundation, the Ministry of Science and Education of Spain, 
the Science and Technology Facilities Council of the United Kingdom, the Higher Education Funding Council for England, the National Center for Supercomputing 
Applications at the University of Illinois at Urbana-Champaign, the Kavli Institute of Cosmological Physics at the University of Chicago, 
the Center for Cosmology and Astro-Particle Physics at the Ohio State University,
the Mitchell Institute for Fundamental Physics and Astronomy at Texas A\&M University, Financiadora de Estudos e Projetos, 
Funda{\c c}{\~a}o Carlos Chagas Filho de Amparo {\`a} Pesquisa do Estado do Rio de Janeiro, Conselho Nacional de Desenvolvimento Cient{\'i}fico e Tecnol{\'o}gico and 
the Minist{\'e}rio da Ci{\^e}ncia, Tecnologia e Inova{\c c}{\~a}o, the Deutsche Forschungsgemeinschaft and the Collaborating Institutions in the Dark Energy Survey. 

The Collaborating Institutions are Argonne National Laboratory, the University of California at Santa Cruz, the University of Cambridge, Centro de Investigaciones Energ{\'e}ticas, 
Medioambientales y Tecnol{\'o}gicas-Madrid, the University of Chicago, University College London, the DES-Brazil Consortium, the University of Edinburgh, 
the Eidgen{\"o}ssische Technische Hochschule (ETH) Z{\"u}rich, 
Fermi National Accelerator Laboratory, the University of Illinois at Urbana-Champaign, the Institut de Ci{\`e}ncies de l'Espai (IEEC/CSIC), 
the Institut de F{\'i}sica d'Altes Energies, Lawrence Berkeley National Laboratory, the Ludwig-Maximilians Universit{\"a}t M{\"u}nchen and the associated Excellence Cluster Universe, 
the University of Michigan, NSF NOIRLab, the University of Nottingham, The Ohio State University, the University of Pennsylvania, the University of Portsmouth, 
SLAC National Accelerator Laboratory, Stanford University, the University of Sussex, Texas A\&M University, and the OzDES Membership Consortium.

Based in part on observations at NSF Cerro Tololo Inter-American Observatory at NSF NOIRLab (NOIRLab Prop. ID 2012B-0001; PI: J. Frieman), which is managed by the Association of Universities for Research in Astronomy (AURA) under a cooperative agreement with the National Science Foundation.

The DES data management system is supported by the National Science Foundation under Grant Numbers AST-1138766 and AST-1536171.
The DES participants from Spanish institutions are partially supported by MICINN under grants PID2021-123012, PID2021-128989 PID2022-141079, SEV-2016-0588, CEX2020-001058-M and CEX2020-001007-S, some of which include ERDF funds from the European Union. IFAE is partially funded by the CERCA program of the Generalitat de Catalunya.

We  acknowledge support from the Brazilian Instituto Nacional de Ci\^encia
e Tecnologia (INCT) do e-Universo (CNPq grant 465376/2014-2).

This manuscript has been authored by Fermi Research Alliance, LLC under Contract No. DE-AC02-07CH11359 with the U.S. Department of Energy, Office of Science, Office of High Energy Physics.
\section*{Data Availability}

Data used in this article is publicly available with the DES5YR data release (Sanchez et al., {\it in prep}).

\section{Appendix}

\begin{figure}
     \centering
     \begin{subfigure}[b]{0.3\textwidth}
         \centering
         \includegraphics[width=\textwidth]{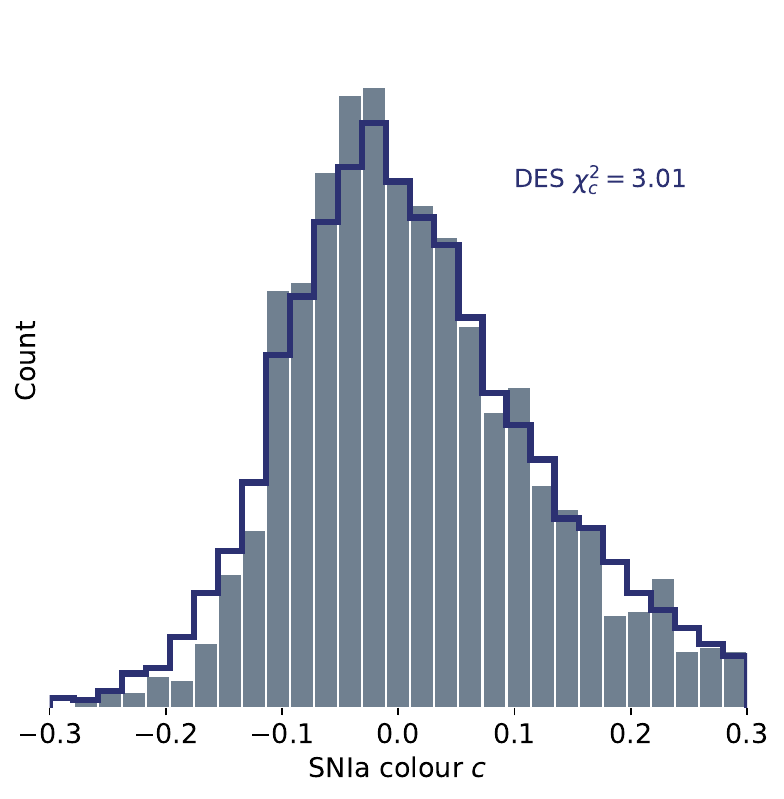}
     \end{subfigure}
     \hfill
     \begin{subfigure}[b]{0.3\textwidth}
         \centering
         \includegraphics[width=\textwidth]{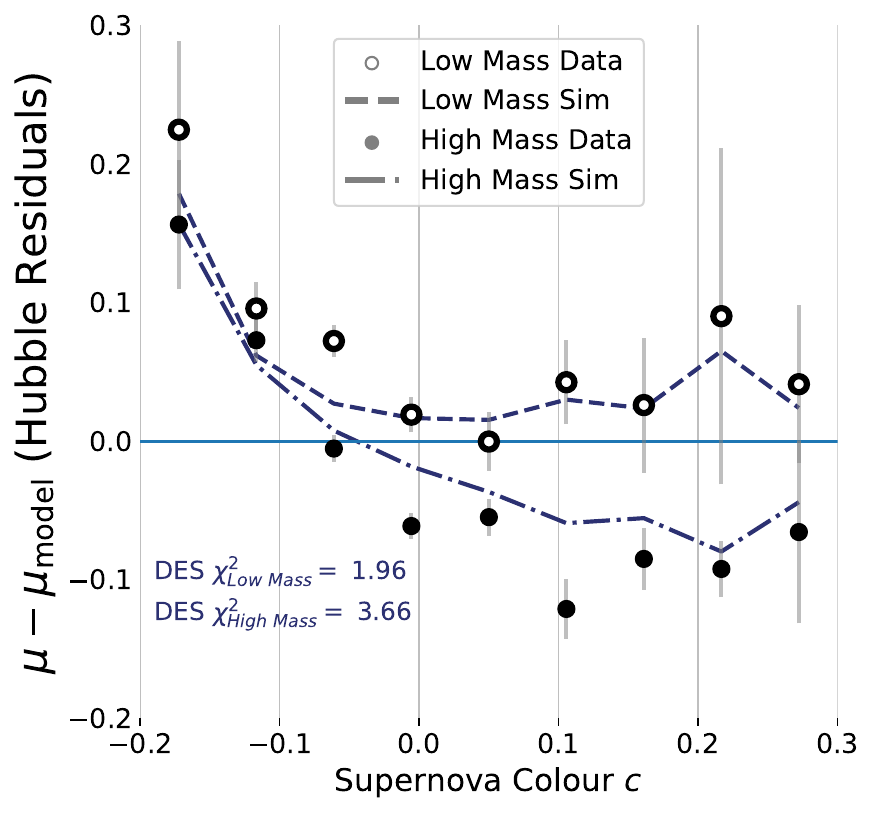}
     \end{subfigure}
     \hfill
     \begin{subfigure}[b]{0.3\textwidth}
         \centering
         \includegraphics[width=\textwidth]{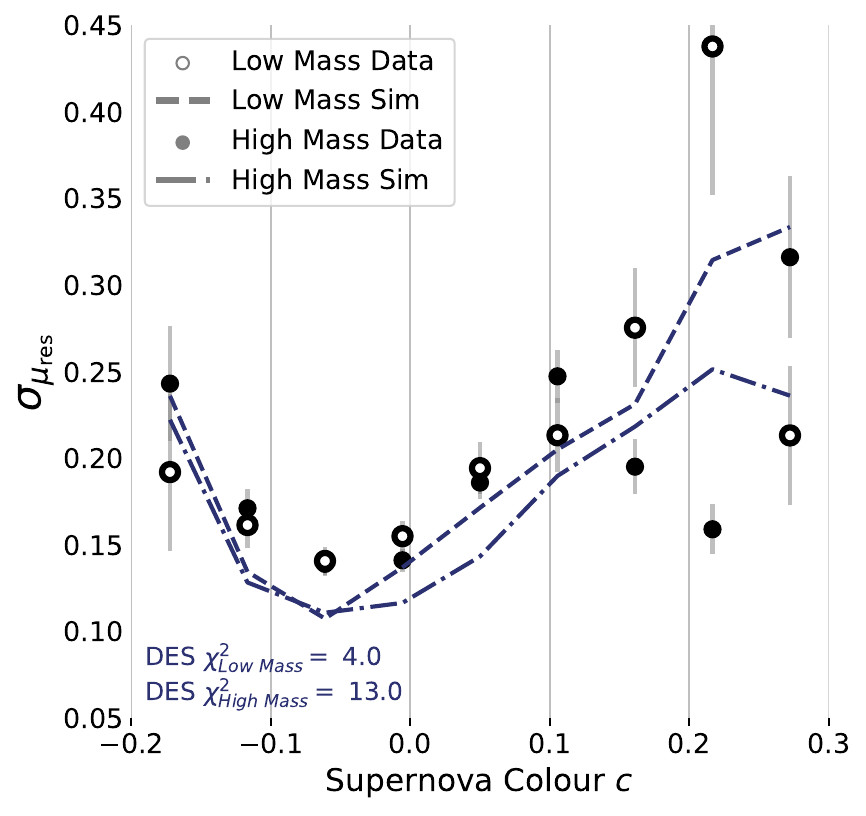}
     \end{subfigure}
        \caption{Goodness-of-fit Criteria for the baseline DES model. Top panel is the colour distribution, with data in grey bars and simulation in solid histogram. Middle panel is the $c$ vs Hubble Residuals relationship, split on the host galaxy stellar mass. High mass mass data is presented in filled circle, low mass data in open circle. High mass simulations are shown in dashed line, low mass simulations in dash-dotted line. Bottom panel shows the scatter in the Hubble Residuals, with the same presentation as the middle panel.}
        \label{fig:DESAPPENDIX}
\end{figure}

\begin{figure}
     \centering
     \begin{subfigure}[b]{0.3\textwidth}
         \centering
         \includegraphics[width=\textwidth]{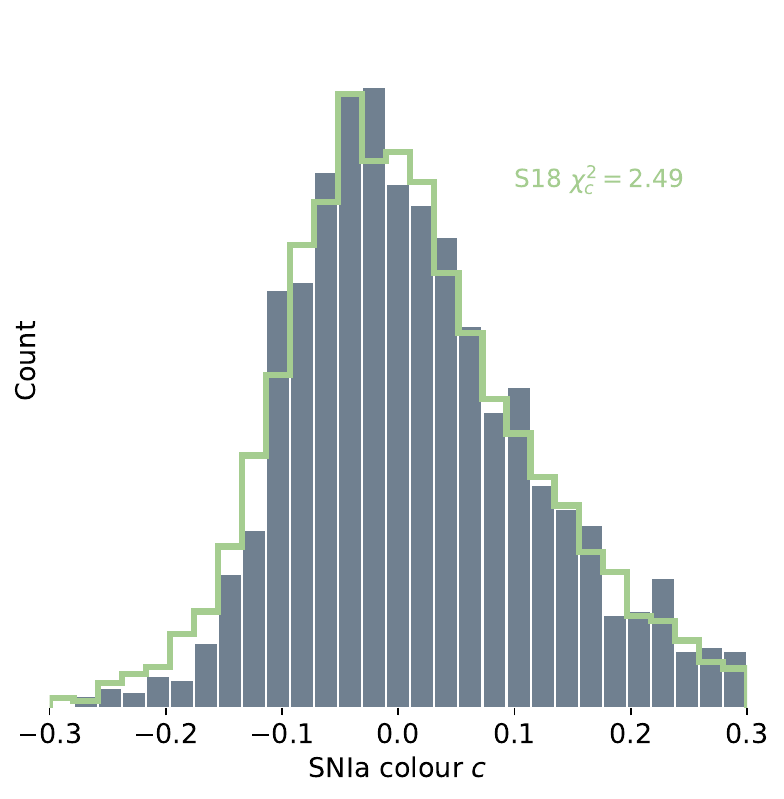}
     \end{subfigure}
     \hfill
     \begin{subfigure}[b]{0.3\textwidth}
         \centering
         \includegraphics[width=\textwidth]{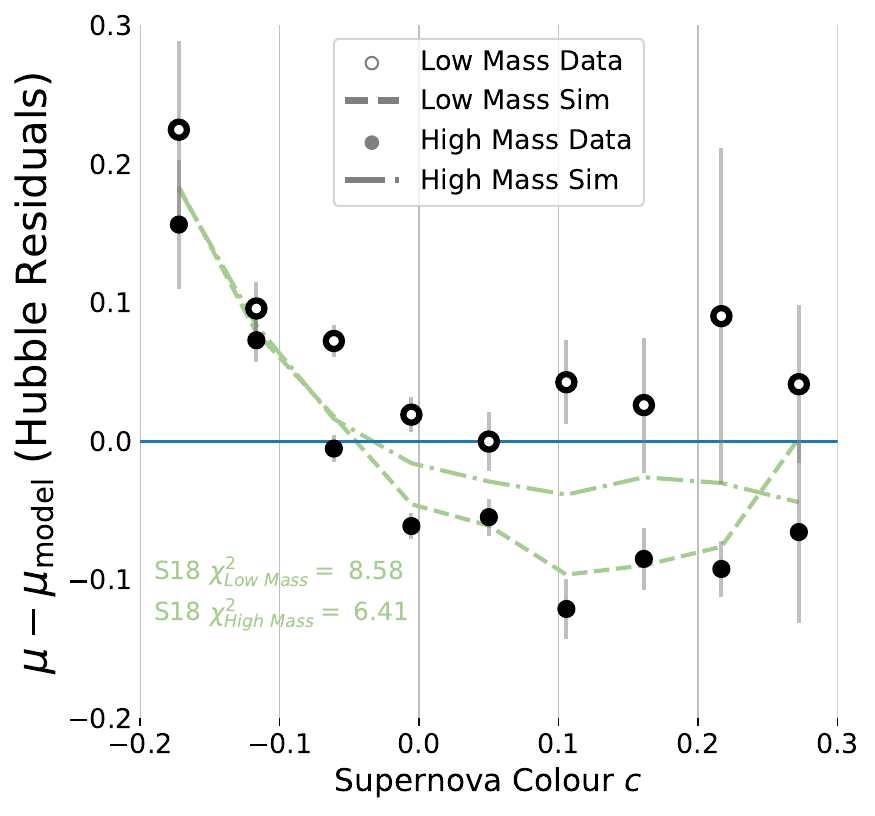}
     \end{subfigure}
     \hfill
     \begin{subfigure}[b]{0.3\textwidth}
         \centering
         \includegraphics[width=\textwidth]{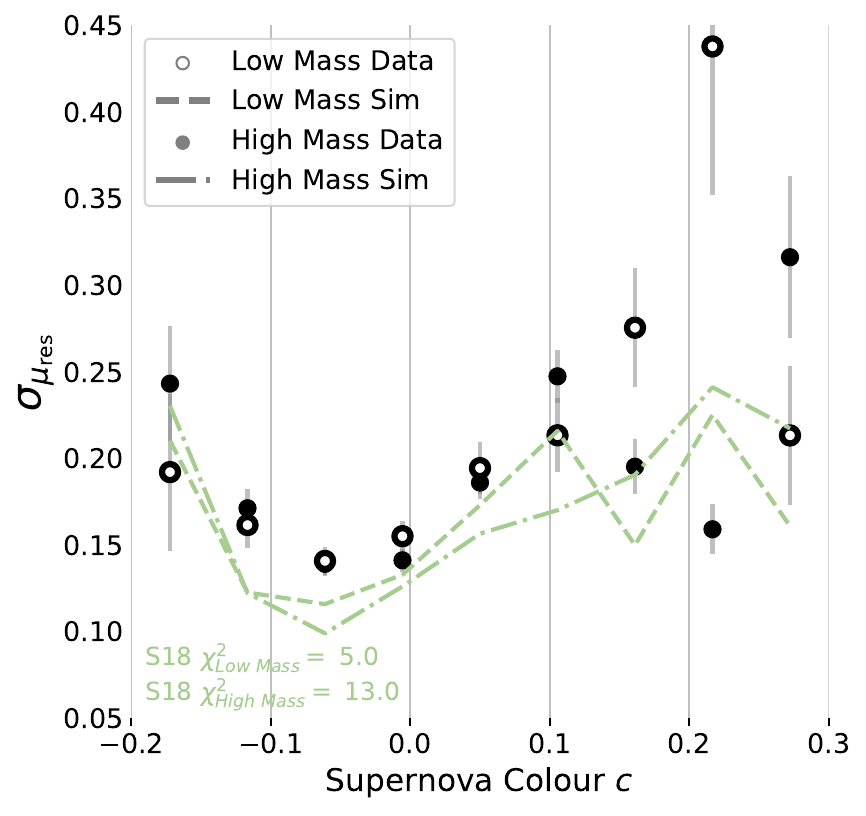}
     \end{subfigure}
        \caption{Goodness-of-fit Criteria for the S18 model. Top panel is the colour distribution, with data in grey bars and simulation in solid histogram. Middle panel is the $c$ vs Hubble Residuals relationship, split on the host galaxy stellar mass. High mass mass data is presented in filled circle, low mass data in open circle. High mass simulations are shown in dashed line, low mass simulations in dash-dotted line. Bottom panel shows the scatter in the Hubble Residuals, with the same presentation as the middle panel.}
        \label{fig:S18APPENDIX}
\end{figure}

\begin{figure}
     \centering
     \begin{subfigure}[b]{0.3\textwidth}
         \centering
         \includegraphics[width=\textwidth]{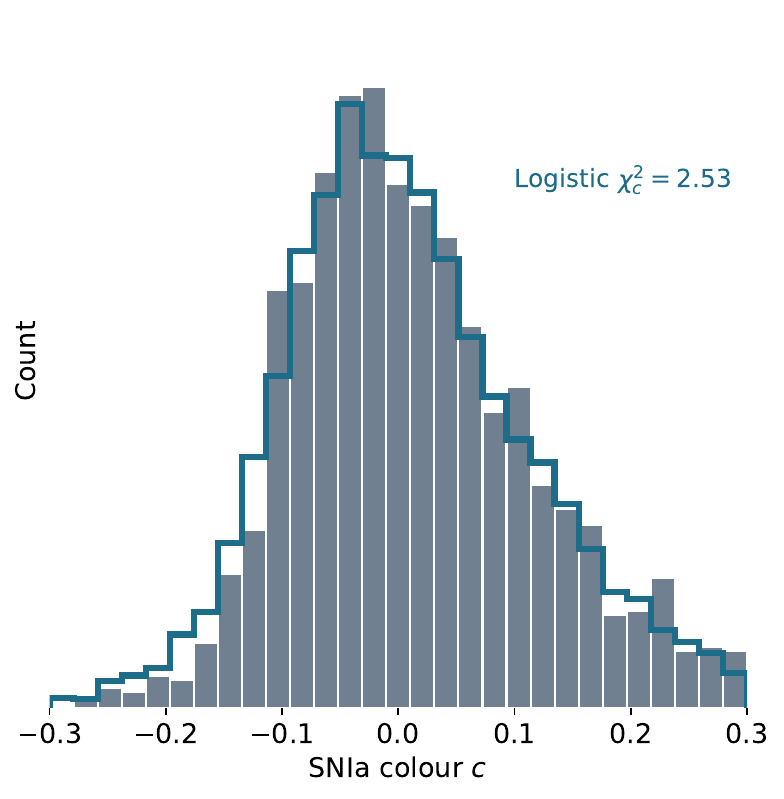}
     \end{subfigure}
     \hfill
     \begin{subfigure}[b]{0.3\textwidth}
         \centering
         \includegraphics[width=\textwidth]{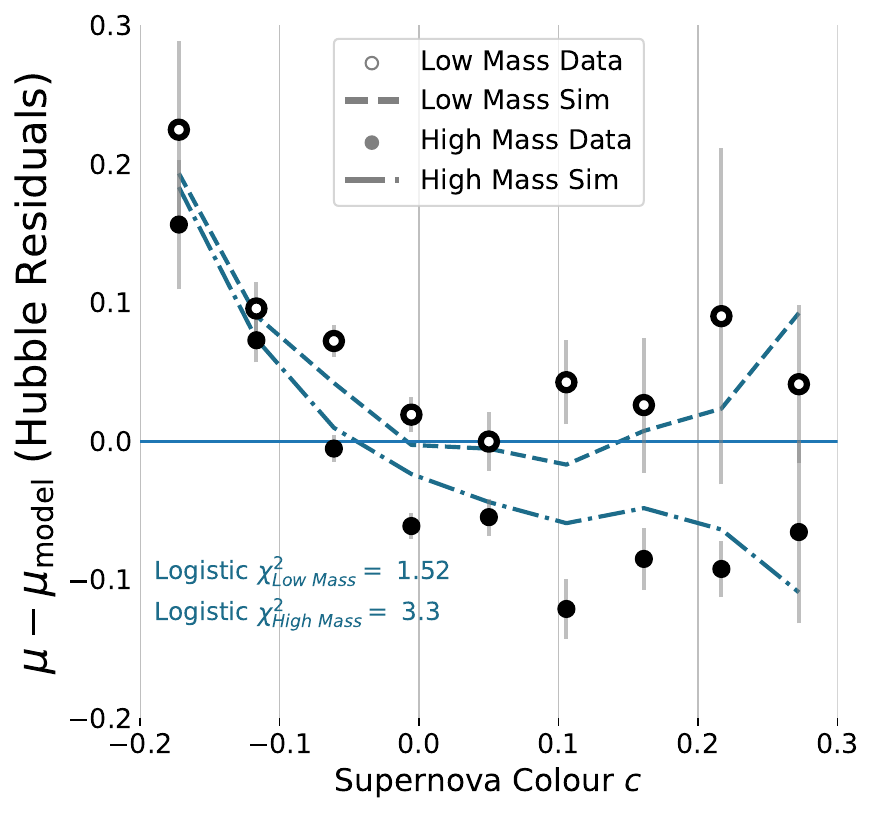}
     \end{subfigure}
     \hfill
     \begin{subfigure}[b]{0.3\textwidth}
         \centering
         \includegraphics[width=\textwidth]{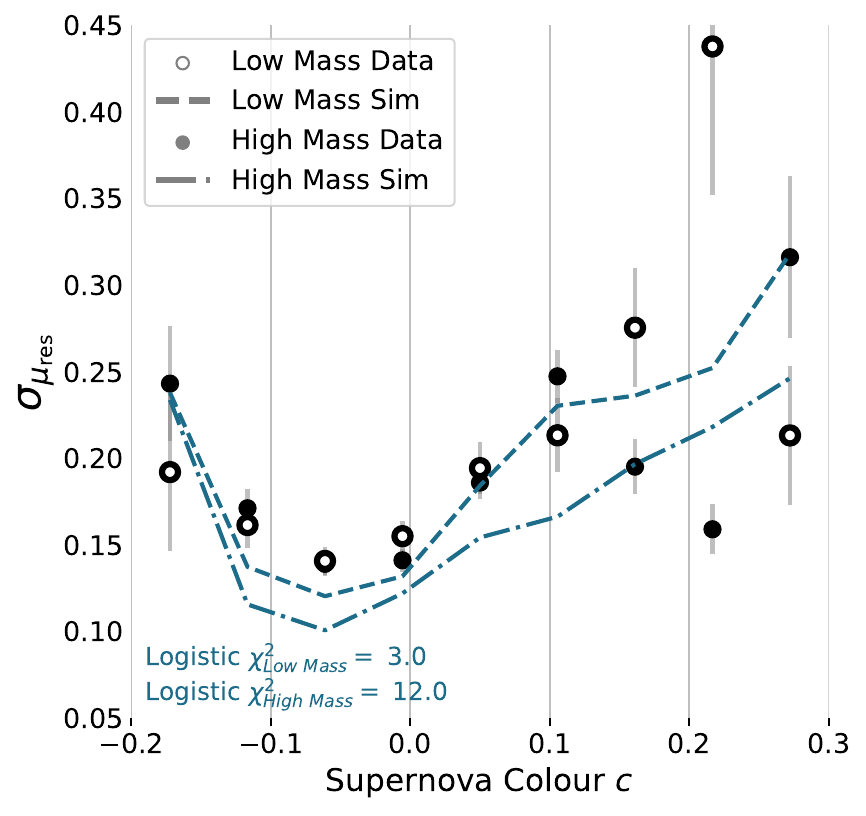}
     \end{subfigure}
        \caption{Goodness-of-fit Criteria for the Logistic model. Top panel is the colour distribution, with data in grey bars and simulation in solid histogram. Middle panel is the $c$ vs Hubble Residuals relationship, split on the host galaxy stellar mass. High mass mass data is presented in filled circle, low mass data in open circle. High mass simulations are shown in dashed line, low mass simulations in dash-dotted line. Bottom panel shows the scatter in the Hubble Residuals, with the same presentation as the middle panel.}
        \label{fig:LOGISTICAPPENDIX}
\end{figure}

\begin{figure}
     \centering
     \begin{subfigure}[b]{0.3\textwidth}
         \centering
         \includegraphics[width=\textwidth]{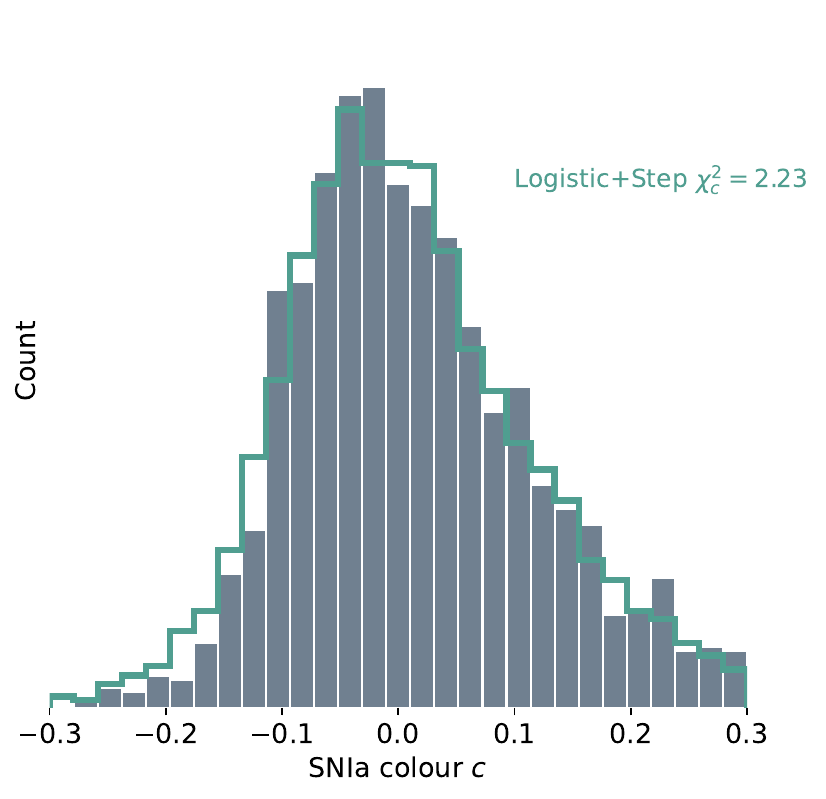}
     \end{subfigure}
     \hfill
     \begin{subfigure}[b]{0.3\textwidth}
         \centering
         \includegraphics[width=\textwidth]{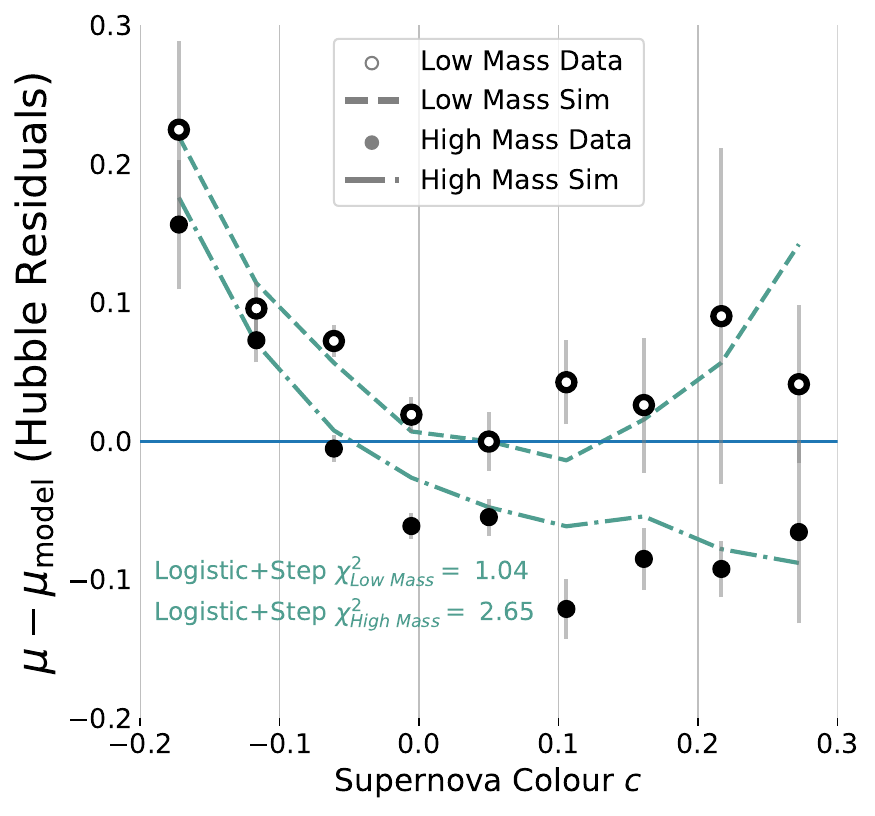}
     \end{subfigure}
     \hfill
     \begin{subfigure}[b]{0.3\textwidth}
         \centering
         \includegraphics[width=\textwidth]{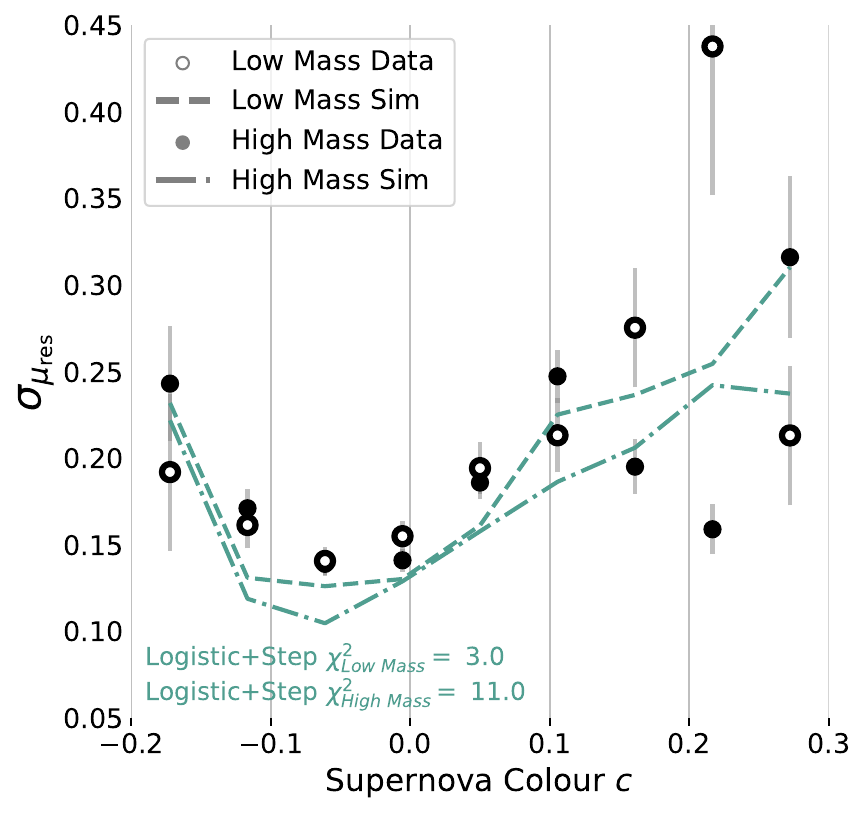}
     \end{subfigure}
        \caption{Goodness-of-fit Criteria for the Logistic+Step model. Top panel is the colour distribution, with data in grey bars and simulation in solid histogram. Middle panel is the $c$ vs Hubble Residuals relationship, split on the host galaxy stellar mass. High mass mass data is presented in filled circle, low mass data in open circle. High mass simulations are shown in dashed line, low mass simulations in dash-dotted line. Bottom panel shows the scatter in the Hubble Residuals, with the same presentation as the middle panel.}
        \label{fig:LOGISTICSTEPAPPENDIX}
\end{figure}

\begin{figure}
     \centering
     \begin{subfigure}[b]{0.3\textwidth}
         \centering
         \includegraphics[width=\textwidth]{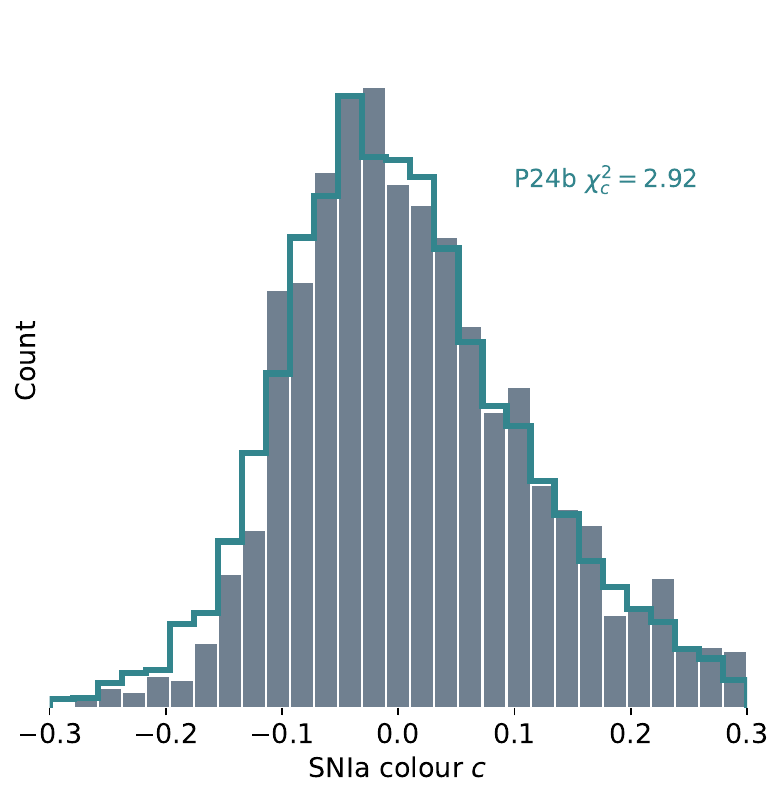}
     \end{subfigure}
     \hfill
     \begin{subfigure}[b]{0.3\textwidth}
         \centering
         \includegraphics[width=\textwidth]{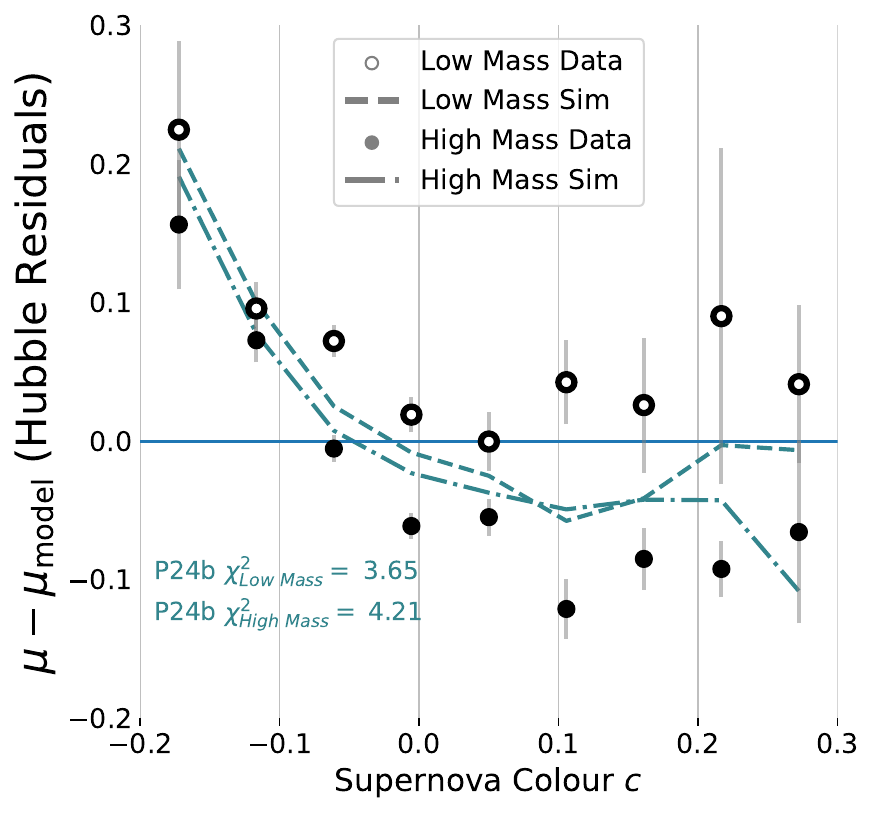}
     \end{subfigure}
     \hfill
     \begin{subfigure}[b]{0.3\textwidth}
         \centering
         \includegraphics[width=\textwidth]{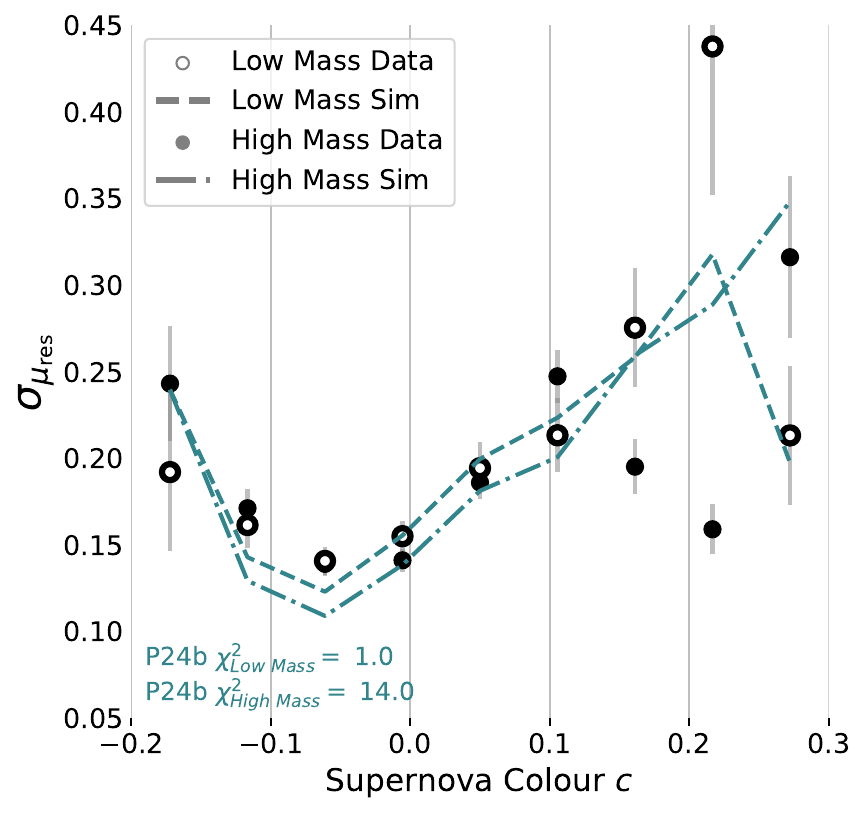}
     \end{subfigure}
        \caption{Goodness-of-fit Criteria for the P24c model. Top panel is the colour distribution, with data in grey bars and simulation in solid histogram. Middle panel is the $c$ vs Hubble Residuals relationship, split on the host galaxy stellar mass. High mass mass data is presented in filled circle, low mass data in open circle. High mass simulations are shown in dashed line, low mass simulations in dash-dotted line. Bottom panel shows the scatter in the Hubble Residuals, with the same presentation as the middle panel.}
        \label{fig:P24CAPPENDIX}
\end{figure}

\begin{figure}
     \centering
     \begin{subfigure}[b]{0.3\textwidth}
         \centering
         \includegraphics[width=\textwidth]{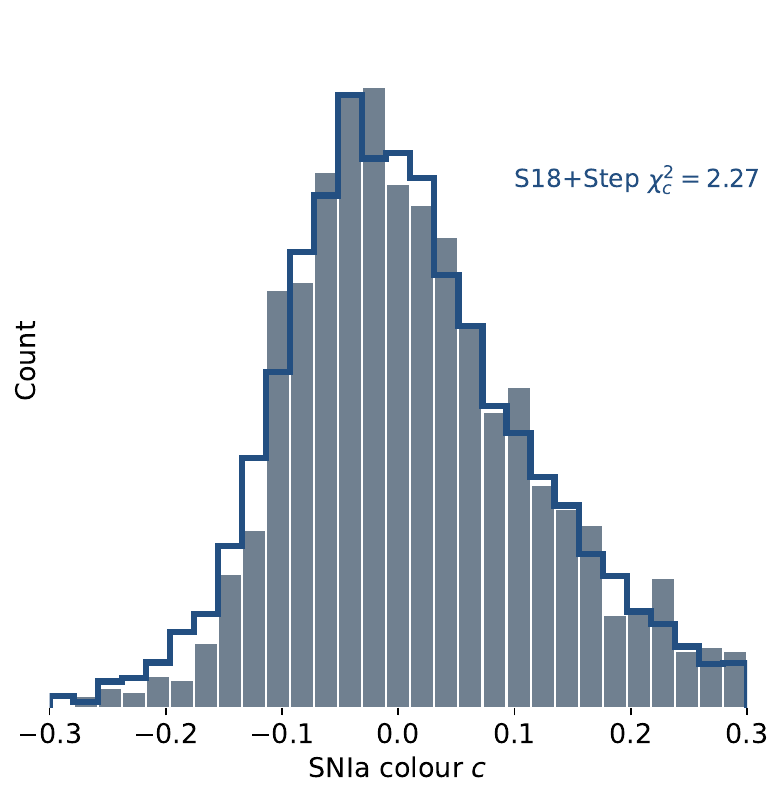}
     \end{subfigure}
     \hfill
     \begin{subfigure}[b]{0.3\textwidth}
         \centering
         \includegraphics[width=\textwidth]{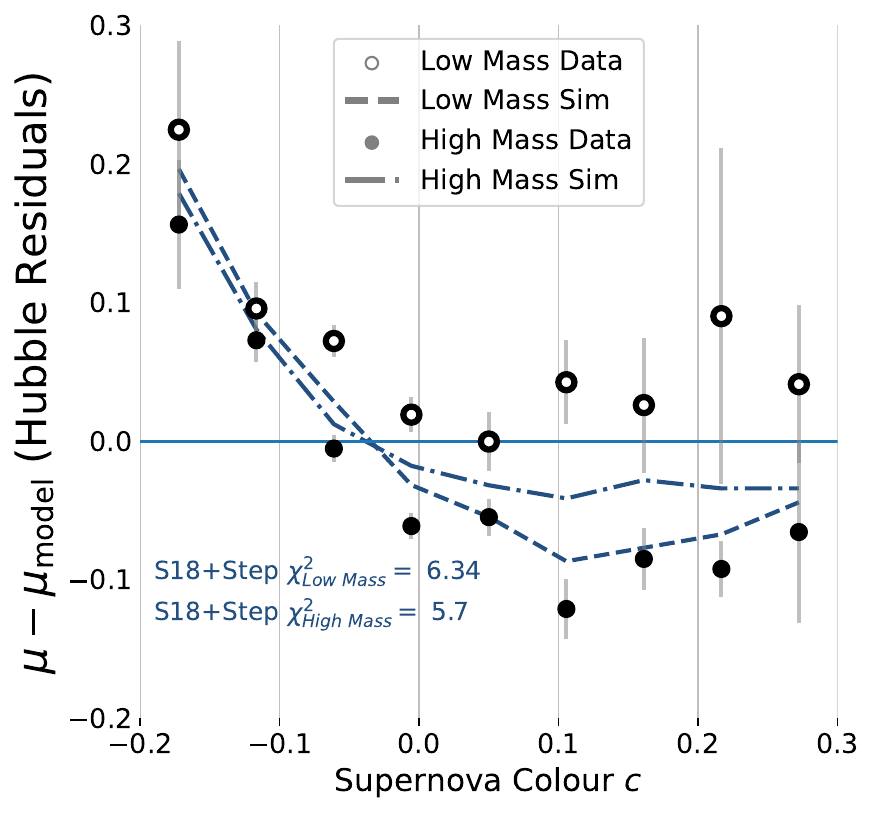}
     \end{subfigure}
     \hfill
     \begin{subfigure}[b]{0.3\textwidth}
         \centering
         \includegraphics[width=\textwidth]{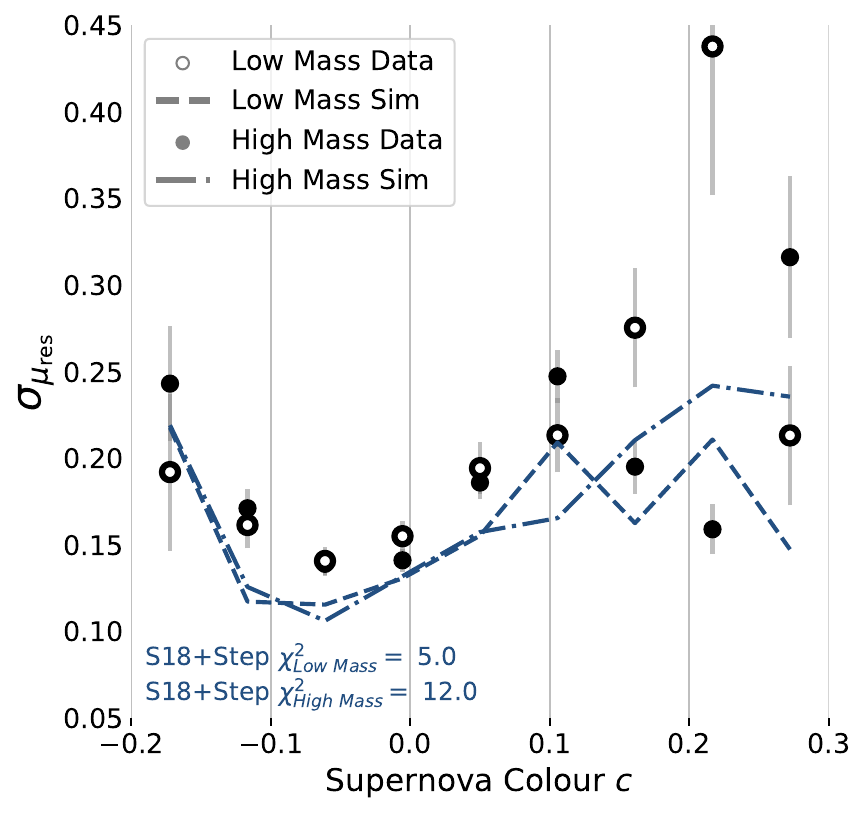}
     \end{subfigure}
        \caption{Goodness-of-fit Criteria for the S18+Step model. Top panel is the colour distribution, with data in grey bars and simulation in solid histogram. Middle panel is the $c$ vs Hubble Residuals relationship, split on the host galaxy stellar mass. High mass mass data is presented in filled circle, low mass data in open circle. High mass simulations are shown in dashed line, low mass simulations in dash-dotted line. Bottom panel shows the scatter in the Hubble Residuals, with the same presentation as the middle panel. }
        \label{fig:S18STEPAPPENDIX}
\end{figure}


\begin{figure}
     \centering
     \begin{subfigure}[b]{0.3\textwidth}
         \centering
         \includegraphics[width=\textwidth]{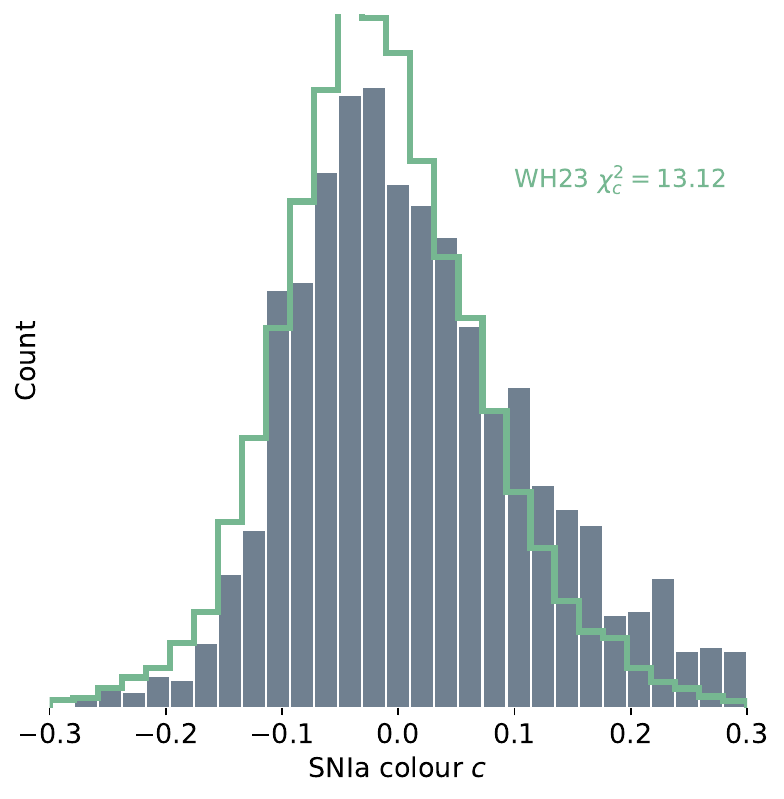}
     \end{subfigure}
     \hfill
     \begin{subfigure}[b]{0.3\textwidth}
         \centering
         \includegraphics[width=\textwidth]{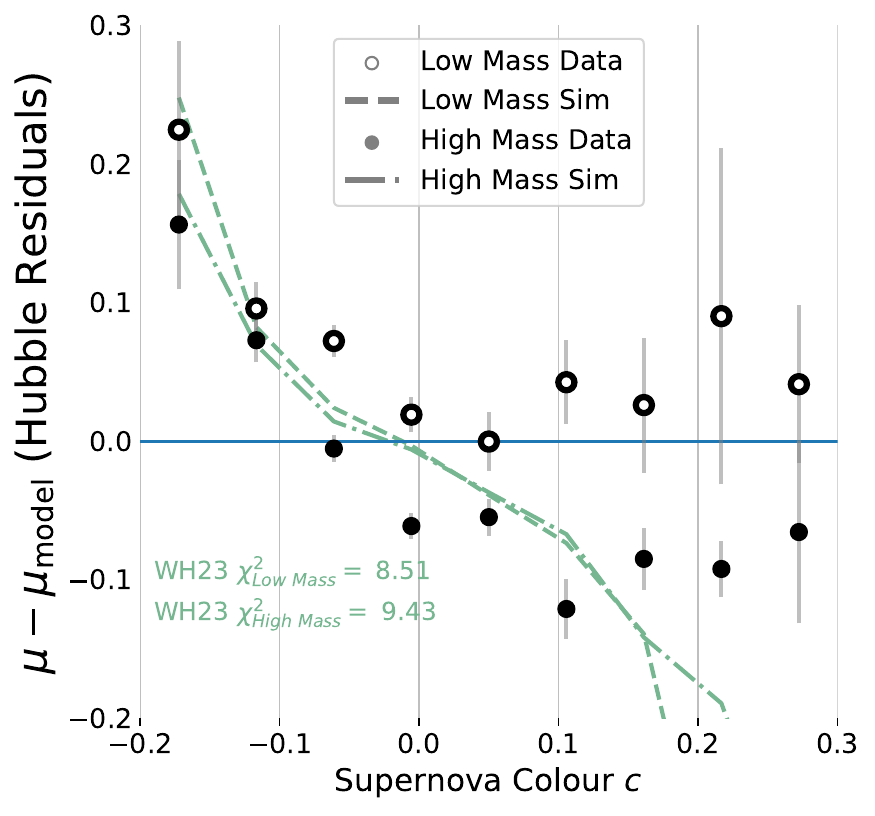}
     \end{subfigure}
     \hfill
     \begin{subfigure}[b]{0.3\textwidth}
         \centering
         \includegraphics[width=\textwidth]{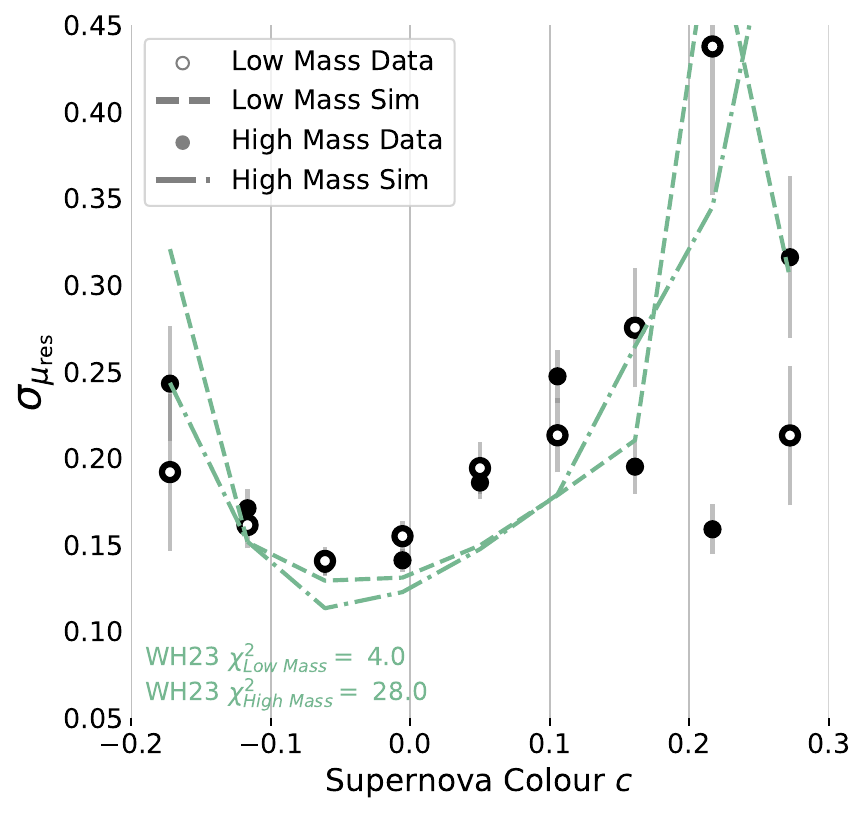}
     \end{subfigure}
        \caption{Goodness-of-fit Criteria for the WH23 model. Top panel is the colour distribution, with data in grey bars and simulation in solid histogram. Middle panel is the $c$ vs Hubble Residuals relationship, split on the host galaxy stellar mass. High mass mass data is presented in filled circle, low mass data in open circle. High mass simulations are shown in dashed line, low mass simulations in dash-dotted line. Bottom panel shows the scatter in the Hubble Residuals, with the same presentation as the middle panel. }
        \label{fig:WH23APPENDIX}
\end{figure}

\section{Affiliations}
$^{1}$ Department of Physics, Duke University Durham, NC 27708, USA\\
$^{2}$ School of Physics and Astronomy, University of Southampton,  Southampton, SO17 1BJ, UK\\
$^{3}$ CENTRA, Instituto Superior T\'ecnico, Universidade de Lisboa, Av. Rovisco Pais 1, 1049-001 Lisboa, Portugal\\
$^{4}$ The Research School of Astronomy and Astrophysics, Australian National University, ACT 2601, Australia\\
$^{5}$ Departamento de Física Teórica and IPARCOS, Universidad Complutense de Madrid, 28040 Madrid, Spain\\
$^{7}$ Center for Astrophysics $\vert$ Harvard \& Smithsonian, 60 Garden Street, Cambridge, MA 02138, USA\\
$^{8}$ INAF-Osservatorio Astronomico di Trieste, via G. B. Tiepolo 11, I-34143 Trieste, Italy\\
$^{9}$ Institut d'Estudis Espacials de Catalunya (IEEC), 08034 Barcelona, Spain\\
$^{10}$ Institute of Space Sciences (ICE, CSIC),  Campus UAB, Carrer de Can Magrans, s/n,  08193 Barcelona, Spain\\
$^{11}$ Centre for Astrophysics \& Supercomputing, Swinburne University of Technology, Victoria 3122, Australia\\
$^{12}$ Institute of Cosmology and Gravitation, University of Portsmouth, Portsmouth, PO1 3FX, UK\\
$^{13}$ Department of Astronomy and Astrophysics, University of Chicago, Chicago, IL 60637, USA\\
$^{14}$ Kavli Institute for Cosmological Physics, University of Chicago, Chicago, IL 60637, USA\\
$^{15}$ Centre for Gravitational Astrophysics, College of Science, The Australian National University, ACT 2601, Australia\\
$^{16}$ Department of Physics and Astronomy, University of Pennsylvania, Philadelphia, PA 19104, USA\\
$^{17}$ Sydney Institute for Astronomy, School of Physics, A28, The University of Sydney, NSW 2006, Australia\\
$^{18}$ School of Mathematics and Physics, University of Surrey, Guildford, Surrey, GU2 7XH, UK\\
$^{19}$ Aix Marseille Univ, CNRS/IN2P3, CPPM, Marseille, France\\
$^{20}$ Cerro Tololo Inter-American Observatory, NSF's National Optical-Infrared Astronomy Research Laboratory, Casilla 603, La Serena, Chile\\
$^{21}$ Laborat\'orio Interinstitucional de e-Astronomia - LIneA, Rua Gal. Jos\'e Cristino 77, Rio de Janeiro, RJ - 20921-400, Brazil\\
$^{22}$ Department of Physics \& Astronomy, University College London, Gower Street, London, WC1E 6BT, UK\\
$^{23}$ Kavli Institute for Particle Astrophysics \& Cosmology, P. O. Box 2450, Stanford University, Stanford, CA 94305, USA\\
$^{24}$ SLAC National Accelerator Laboratory, Menlo Park, CA 94025, USA\\
$^{25}$ Instituto de Astrofisica de Canarias, E-38205 La Laguna, Tenerife, Spain\\
$^{26}$ Institut de F\'{\i}sica d'Altes Energies (IFAE), The Barcelona Institute of Science and Technology, Campus UAB, 08193 Bellaterra (Barcelona) Spain\\
$^{27}$ Hamburger Sternwarte, Universit\"{a}t Hamburg, Gojenbergsweg 112, 21029 Hamburg, Germany\\
$^{28}$ School of Mathematics and Physics, University of Queensland,  Brisbane, QLD 4072, Australia\\
$^{29}$ Department of Physics, IIT Hyderabad, Kandi, Telangana 502285, India\\
$^{30}$ Jet Propulsion Laboratory, California Institute of Technology, 4800 Oak Grove Dr., Pasadena, CA 91109, USA\\
$^{31}$ Institute of Theoretical Astrophysics, University of Oslo. P.O. Box 1029 Blindern, NO-0315 Oslo, Norway\\
$^{32}$ Fermi National Accelerator Laboratory, P. O. Box 500, Batavia, IL 60510, USA\\
$^{33}$ Instituto de Fisica Teorica UAM/CSIC, Universidad Autonoma de Madrid, 28049 Madrid, Spain\\
$^{34}$ Center for Astrophysical Surveys, National Center for Supercomputing Applications, 1205 West Clark St., Urbana, IL 61801, USA\\
$^{35}$ Department of Astronomy, University of Illinois at Urbana-Champaign, 1002 W. Green Street, Urbana, IL 61801, USA\\
$^{36}$ Santa Cruz Institute for Particle Physics, Santa Cruz, CA 95064, USA\\
$^{37}$ Center for Cosmology and Astro-Particle Physics, The Ohio State University, Columbus, OH 43210, USA\\
$^{38}$ Department of Physics, The Ohio State University, Columbus, OH 43210, USA\\
$^{39}$ Australian Astronomical Optics, Macquarie University, North Ryde, NSW 2113, Australia\\
$^{40}$ Lowell Observatory, 1400 Mars Hill Rd, Flagstaff, AZ 86001, USA\\
$^{41}$ George P. and Cynthia Woods Mitchell Institute for Fundamental Physics and Astronomy, and Department of Physics and Astronomy, Texas A\&M University, College Station, TX 77843,  USA\\
$^{42}$ LPSC Grenoble - 53, Avenue des Martyrs 38026 Grenoble, France\\
$^{43}$ Instituci\'o Catalana de Recerca i Estudis Avan\c{c}ats, E-08010 Barcelona, Spain\\
$^{44}$ Department of Astrophysical Sciences, Princeton University, Peyton Hall, Princeton, NJ 08544, USA\\
$^{45}$ Observat\'orio Nacional, Rua Gal. Jos\'e Cristino 77, Rio de Janeiro, RJ - 20921-400, Brazil\\
$^{46}$ Department of Physics, Carnegie Mellon University, Pittsburgh, Pennsylvania 15312, USA\\
$^{47}$ Centro de Investigaciones Energ\'eticas, Medioambientales y Tecnol\'ogicas (CIEMAT), Madrid, Spain\\
$^{48}$ Department of Physics, University of Michigan, Ann Arbor, MI 48109, USA\\
$^{49}$ Computer Science and Mathematics Division, Oak Ridge National Laboratory, Oak Ridge, TN 37831\\
$^{50}$ Department of Astronomy, University of California, Berkeley,  501 Campbell Hall, Berkeley, CA 94720, USA\\
$^{51}$ Lawrence Berkeley National Laboratory, 1 Cyclotron Road, Berkeley, CA 94720, USA\\



\bibliographystyle{mnras}
\bibliography{research2} 




\bsp	
\label{lastpage}
\end{document}